\documentclass[article,nojss,shortnames]{jss}

\usepackage{thumbpdf}
\usepackage{amsfonts,amstext,amsmath,amssymb,amsthm, mathtools}
\usepackage{accents}
\usepackage{rotating}
\usepackage{verbatim}
\usepackage{booktabs}
\usepackage{makecell}
\usepackage{caption}
\usepackage{multirow}
\usepackage{relsize}
\usepackage{diagbox}
\usepackage{numprint}
\usepackage{dcolumn}
\usepackage{pifont}
\usepackage{alltt}
\usepackage{xcolor}

\usepackage[]{graphicx}\usepackage[]{xcolor}
\makeatletter
\def\maxwidth{ \ifdim\Gin@nat@width>\linewidth
	\linewidth
	\else
	\Gin@nat@width
	\fi
}
\makeatother

\definecolor{fgcolor}{rgb}{0.345, 0.345, 0.345}
\newcommand{\hlnum}[1]{\textcolor[rgb]{0.686,0.059,0.569}{#1}}\newcommand{\hlstr}[1]{\textcolor[rgb]{0.192,0.494,0.8}{#1}}\newcommand{\hlcom}[1]{\textcolor[rgb]{0.678,0.584,0.686}{\textit{#1}}}\newcommand{\hlopt}[1]{\textcolor[rgb]{0,0,0}{#1}}\newcommand{\hlstd}[1]{\textcolor[rgb]{0.345,0.345,0.345}{#1}}\newcommand{\hlkwb}[1]{\textcolor[rgb]{0.69,0.353,0.396}{#1}}\newcommand{\hlkwc}[1]{\textcolor[rgb]{0.333,0.667,0.333}{#1}}\newcommand{\hlkwd}[1]{\textcolor[rgb]{0.737,0.353,0.396}{\textbf{#1}}}
\definecolor{shadecolor}{rgb}{.97, .97, .97}
\definecolor{messagecolor}{rgb}{0, 0, 0}
\definecolor{warningcolor}{rgb}{1, 0, 1}
\definecolor{errorcolor}{rgb}{1, 0, 0}
\newenvironment{knitrout}{}{}

\usepackage{framed}
\makeatletter
\newenvironment{kframe}{\def\at@end@of@kframe{}\ifinner\ifhmode \def\at@end@of@kframe{\end{minipage}}\begin{minipage}{\columnwidth}\fi\fi \def\FrameCommand##1{\hskip\@totalleftmargin \hskip-\fboxsep
		\colorbox{shadecolor}{##1}\hskip-\fboxsep
\hskip-\linewidth \hskip-\@totalleftmargin \hskip\columnwidth}\MakeFramed {\advance\hsize-\width
		\@totalleftmargin\z@ \linewidth\hsize
		\@setminipage}}{\par\unskip\endMakeFramed \at@end@of@kframe}
\makeatother

\newcommand{\Darrow}{\stackrel{D}{\longrightarrow}}

\newcommand{\Sens}{\text{Sens}}
\newcommand{\Spec}{\text{Spec}}
\newcommand{\AUC}{\text{AUC}}
\newcommand{\ROC}{\text{ROC}}
\newcommand{\OVL}{\text{OVL}}
\newcommand{\LOD}{\text{LOD}}

\newcommand{\FZ}{F_Z}
\newcommand{\FZinv}{F_Z^{-1}}

\newcommand{\FNinv}{F_0^{-1}}
\newcommand{\hinv}{h^{-1}}

\newcommand{\rX}{\mX}

\newcommand{\ry}{y}

\newcommand{\rx}{\xvec}

\newcommand{\h}{h}

\newcommand{\basisy}{\bvec}

\newcommand{\parm}{\varthetavec}

\newcommand{\hparm}{\varthetahatvec}

\newcommand{\shiftparm}{\betavec}
\newcommand{\eshiftparm}{\delta}

\newcommand{\hshiftparm}{\betahatvec}
\newcommand{\absshift}{|\eshiftparm|}
\newcommand{\intparm}{\gammavec}
\newcommand{\eintparm}{\gamma}
\newcommand{\xparm}{\xivec}
\newcommand{\exparm}{\xi}

\newcommand{\V}{\mathbb{V}}
\newcommand{\RR}{\mathbb{R}}

\usepackage{dsfont}

 \DeclareMathOperator{\logit}{logit}
 \DeclareMathOperator{\probit}{probit}
 \DeclareMathOperator{\expit}{expit}
 \DeclareMathOperator{\loglog}{loglog}
 \DeclareMathOperator{\cloglog}{cloglog}

 \DeclareMathOperator*{\argmax}{{arg\,max}}

\def \bvec {\text{\boldmath$b$}}

\def \xvec {\text{\boldmath$x$}}    \def \mX {\text{\boldmath$X$}}
\def \yvec {\text{\boldmath$y$}}    \def \mY {\text{\boldmath$Y$}}
    \def \mZ {\text{\boldmath$Z$}}

\def \hhat {\text{$\hat h$}}

\def \eshiftparmtilde {\text{$\tilde \eshiftparm$}}

\def \betavec         {\text{\boldmath$\beta$}}
\def \gammavec        {\text{\boldmath$\gamma$}}

\def \thetavec        {\text{\boldmath$\theta$}}
\def \varthetavec     {\text{\boldmath$\vartheta$}}

\def \xivec           {\text{\boldmath$\xi$}}

\def \varsigmavec     {\text{\boldmath$\varsigma$}}

\def \betahatvec         {\text{\boldmath$\hat \beta$}}

\def \varthetahatvec     {\text{\boldmath$\hat \vartheta$}}

\def \gammatildevec        {\text{\boldmath$\tilde \gamma$}}

\def \thetatildevec        {\text{\boldmath$\tilde \theta$}}

\def \xitildevec           {\text{\boldmath$\tilde \xi$}}

\def \mLambda  {\mathbf{\Lambda}}

\def \mSigma   {\mathbf{\Sigma}}

\def \mhatTheta   {\mathbf{\hat \Theta}}

\def \nullvec {\mathbf{0}}

\newcommand{\cmark}{\ding{51}}%

\newcommand{\AYcite}[2]{\cite{#2}}
\newcommand{\SMref}[2]{\ref{#2}}

\author{Ainesh Sewak \\ Universit\"at Z\"urich
	\And Torsten Hothorn \\ Universit\"at Z\"urich}
\Plainauthor{Ainesh Sewak, Torsten Hothorn}

\title{Transformation models for ROC analysis}
\Plaintitle{}

\Abstract{
	Receiver operating characteristic (ROC) analysis is one of the most popular approaches for evaluating and comparing the accuracy of medical diagnostic tests. 
Although various methodologies have been developed for estimating ROC curves and its associated summary indices, there is no consensus on a single framework that can provide consistent statistical inference whilst handling the complexities associated with medical data.
Such complexities might include covariates that influence the diagnostic potential of a test, ordinal test data, censored data due to instrument detection limits or correlated biomarkers.
We propose a regression model for the transformed test results which exploits the invariance of ROC curves to monotonic transformations and naturally accommodates these features.
Our use of maximum likelihood inference guarantees asymptotic efficiency of the resulting estimators and associated confidence intervals.
Simulation studies show that the estimates based on transformation models are unbiased and yield coverage at nominal levels.
The methodology is applied to a cross-sectional study of metabolic syndrome where we investigate the covariate-specific performance of weight-to-height ratio as a non-invasive diagnostic test.  
Software implementations for all the methods described in the article are provided in the \pkg{tram} \proglang{R} package.
 }

\Plainkeywords{transformation model, ROC curve, AUC, diagnostic tests, distribution regression, ordinal outcome, censoring, Youden Index, OVL, correlated biomarkers, limit of detection}

\Address{
  Ainesh Sewak, Torsten Hothorn\\
  Institut f\"ur Epidemiologie, Biostatistik und Pr\"avention \\
  Universit\"at Z\"urich \\
  Hirschengraben 84, CH-8001 Z\"urich, Switzerland \\
  \texttt{Ainesh.Sewak@uzh.ch}, \texttt{Torsten.Hothorn@R-project.org} \\

}

\begin{document}

\section{Introduction}
\label{sec:intro}

Estimating receiver operating characteristic (ROC) curves for evaluating the performance of medical diagnostic tests has been a main focus of statistical literature over the last decades \citep{green1966signal, egan1975signal, pepe2003statistical, zou2011statistical, inacio2021statistical}. Diagnostic tests screen for the presence or absence of a disease. Characterizing their accuracy is essential to ensure appropriate prevention, treatment and monitoring of diseases. ROC curves are a valuable tool in determining the diagnostic potential of a test and continue to be extensively applied in biomedical studies as new tests or biomarkers are developed in radiology, oncology, genetics, and other related fields.
Increasingly more applications can be expected due to advancements in technology and analyzing resulting data requires a computationally straightforward approach to provide accurate and consistent statistical inference.

Previous research has focused on extending statistical methodology for ROC curve estimation to address issues such as adjustment for covariates \citep{pepe1997regression, pepe1998three, pepe2000interpretation, faraggi2003adjusting}, incorporating censoring due to instrument detection limits~\citep{perkins2007receiver, ruopp2008youden, bantis2017estimation}, comparing correlated diagnostic tests~\citep{hanley1983method, delong1988comparing, zou2002semiparametric, molodianovitch2006comparing, bantis2016comparison} and robustness to model misspecification~\citep{bianco2020robust, inacio2021robust}. In addition, a wide variety of parametric and nonparametric methods have been proposed within frequentist and Bayesian paradigms \citep{hsieh1996nonparametric, alonzo2002distribution, erkanli2006bayesian, branscum2008bayesian, yao2010nonparametric, gonzalez2011roc, rodriguez2011roc, cai2002semiparametric, cai2004semi, inacio2013bayesian, ghosal2022estimation}. However, there is no consensus on an analytic approach that can handle all these issues simultaneously.

An attractive feature of the ROC curve, which has scantly been used for its estimation, is that it remains invariant to monotonic transformations of the test results. Although transformations have been used to bring continuous test results into a form that approximately satisfies the assumptions of a suitable parametric model~\citep{faraggi2003adjusting, schisterman2004adjusting}, estimation of a transformation function has been limited to the Box-Cox power transformation family~\citep{box1964analysis, zou1998original, zou2000two}. For rank-based methods, the transformation function can be left unspecified but in all cases a restriction to normality has been previously imposed on the model for the ROC curve~\citep{zou1997smooth}.

In this article we present a new unifying methodological framework for estimating ROC curves and its associated summary indices by modeling the relationship between the \emph{transformed} test results and potential covariates. We employ transformation models to jointly estimate the transformation function and regression parameters~\citep{hothorn2014conditional, hothorn2018most}. This approach specifies a parametric model for the ROC curve but remains distribution-free because we do not impose any strong assumptions on the transformation function. Using the estimated parameters, we show how to evaluate covariate effects on discriminatory performance of diagnostic tests. We focus on data obtained from case-control or cross-sectional studies of subjects with continuous or ordinal test results. Unlike nonparametric methods which are flexible but difficult to interpret and implement, transformation models excel on both fronts. Their output can be easily interpreted by clinicians, which is of important consideration in biomedical applications. \proglang{R}~implementations of all methods discussed in this article are available, along with a set of supporting examples. The goals of this article are to introduce a general methodology for performing ROC analysis which is model-based but distribution-free; to apply maximum likelihood inference procedures to examine covariate effects with potentially censored observations; and to demonstrate methods for testing differences between correlated ROC curves of several tests at specific covariate levels within the same conceptual and computational framework.

\subsection{Notation}

Let the random variable $Y$ denote the continuous result of a diagnostic test and let $D$ denote the disease status, with $D=1$ if
a subject is diseased and 0 if nondiseased. We denote quantities
conditional on the disease status using subscripts. For example, $Y_1$ and $Y_0$ are the test results in the diseased and nondiseased populations with cumulative distribution functions (CDF) given by $F_1$~and~$F_0$ and densities $f_1$~and~$f_0$, respectively. Suppose that the subject is diagnosed as diseased when their test result exceeds a threshold value,~$c$. By convention, we assume that larger values of the test result are more indicative of the disease. The probability of truly identifying a diseased and nondiseased subject are defined as sensitivity, $\Prob(Y_1 > c)=1-F_1(c)$, and specificity, $\Prob(Y_0 \leq c)=F_0(c)$, respectively. The set of pairs $(1-\text{specificity}, \text{sensitivity})$ for all $c \in \RR$ produce the ROC curve. By setting~$p=1-F_0(c)$, an equivalent representation of the ROC curve is
\begin{align*}
\ROC(p) = 1 - F_1 \left( \FNinv(1-p) \right).
\end{align*}

\subsection{Summary indices}
Summary indices of the ROC curve quantify the degree of separation between the distributions $Y_1$~and~$Y_0$. The most widely used index is the area under the ROC curve (AUC) defined by
\begin{align*}
	\AUC = \Prob(Y_1 > Y_0) = \int_0^1 \ROC(p) dp.
\end{align*}
The AUC represents the probability that the test results of a randomly selected diseased subject exceeds the one of a nondiseased subject and is directly related to the Mann-Whitney-Wilcoxon U-statistic (MWW) \citep{bamber1975area, hanley1982meaning}. 
Alternative indices include the Youden index \citep{youden1950index}, $J$,  which combines sensitivity and specificity over all possible thresholds to provide the maximum potential effectiveness of a diagnostic test, given by 
\begin{align*}
	J &=\max_{c \in \RR} [F_0(c) - F_1(c)].
\end{align*}
The Youden index is equivalent to the Smirnov (or the two-sample Kolmogorov-Smirnov) test statistic \citep{gail1976generalization, atsushi2022} and can be represented as half the $L_1$ distance between two densities or as the complement of the overlapping coefficient (OVL) \citep{weitzman1970measures, feller1991intro,schmid2006nonparametric, martinez2022use}:
\begin{align*}
	J &= \frac{1}{2} \int |f_0(y) - f_1(y)| dy= 1- \int \min[f_1(y), f_0(y)] dy =1 - \OVL.
\end{align*}
Additionally, the threshold corresponding to $J$, where sensitivity and specificity are maximized, denoted as $c^*$, is often used in clinical practice as the optimal classification threshold to screen subjects.

\subsection{Covariates}
The result of a diagnostic test can be affected by both the disease status, as well as by covariates associated with the subjects. Consequently, the test result distributions and their separation may differ by levels of the covariates. In order to appropriately
understand the accuracy of the test in subpopulations, we can use
covariate-specific or conditional ROC curves \citep{tosteson1988general,
	pepe1997regression}. Throughout we use boldface letters to denote vectors and
lightface for scalars. Let $\rX$ denote a vector of covariates that are hypothesized to have an impact on the accuracy of the test. The conditional CDF in the diseased population is given by $F_1(y \mid \rx) = \Prob(Y_1 \leq y \mid
\rX = \rx)$ and analogously given for the nondiseased population. The
covariate-specific ROC can be written as
\begin{align}
\label{eq:rocx}
	\ROC(p \mid \rx) = 1 - F_1 \left( \FNinv(1-p \mid \rx) \mid \rx \right)
\end{align}
with its counterpart conditional summary indices, $\AUC(\rx)$ and $J(\rx)$,
defined accordingly.

The covariate-specific ROC curve can be generated by modeling the conditional distribution of the test results. This is known as the induced or indirect methodology for modeling ROC curves \citep{pepe1997regression, pepe1998three, pepe2003statistical, faraggi2003adjusting, rodriguez2011roc}. Alternatively, direct methodology is based on a regression model for the ROC curve itself \citep{pepe2000interpretation, alonzo2002distribution, cai2004semi}.

\subsection{Overview}
The article proceeds as follows. In Section~\ref{sec:methods}, we propose a transformation modeling framework for parameterizing ROC curves from which we derive closed-form expressions for associated AUC and Youden summary indices. We then show how our approach addresses the main components of ROC analysis: summarizing the discriminatory capabilities of a diagnostic test for two-sample classification, adjusting for covariates, and comparing the efficacy of two (or more) correlated diagnostic tests. We discuss maximum likelihood estimation procedures for our model and corresponding inference. Specifically, we develop score confidence bands for the ROC curve and score confidence intervals for its summary indices, which remain invariant to reparametrizations and hence can be determined from the estimated parameters of the transformation model.
In Section~\ref{sec:application} we apply our approach to a cross-sectional study for detection of metabolic syndrome. Specifically, using the weight-to-height ratio as a non-invasive diagnostic test for metabolic syndrome, we investigate the age and gender-specific performance of the test. We conclude the article with a discussion.

\section{Methods}
\label{sec:methods}

\subsection{Transformation model}
\label{sec:trafo}

The ROC curve is a composition of distribution functions and thus is invariant to strictly monotonically increasing transformations of $Y$. We propose a model for the conditional distribution of the transformed test result given the disease status and covariates. This transformation is obtained from the data and leads to a distribution-free framework to parameterize the covariate-specific ROC curve and its summary indices.

Suppose there exists a strictly monotonically increasing function $h$ such that
the relationship between the transformed test result and the covariates
follows a shift-scale model
\begin{align*}
	h(Y) &= \mu_d(\rx) + \sigma_d(\rx) Z,
\end{align*}
where $D=d$ specifies the disease indicator ($D=0$ for nondiseased and $D=1$ for diseased), $\rX=\rx$ a fixed set of covariates, $\mu_d(\rx)$ the shift term, $\sigma_d(\rx)$ the scale term and~$Z \in \RR$ is a latent random variable with an \emph{a priori} known absolutely continuous log-concave CDF, $\FZ$.
Given that $D$ and $\rX$ are fixed, the conditional
CDF for $Y$ is
\begin{align}
\begin{split}
	\Prob(Y \leq y \mid D=d, \rX = \rx) &= F_d(y \mid \rx)  = \FZ \left(  \frac{h(y) - \mu_d(\rx)}{\sigma_d(\rx)} \right).
	\label{eq:cdf_gen}
\end{split}
\end{align}
Equation~\ref{eq:cdf_gen} represents a general class of models called
transformation models \citep{bickel1981analysis, bickel1986efficient,
bickel1993efficient, hothorn2018most}. The transformation function~$h$ uniquely characterizes the distribution of~$Y$, similar to the density or distribution function.
Plugging in this conditional CDF of~$Y$ into Equation~\ref{eq:rocx}, $h$ cancels out and the covariate-specific ROC curve is given by
\begin{align}
\begin{split}
	\ROC(p \mid \rx ) &= 1-\FZ \left( \zeta(\rx) \FZinv(1-p) - \eshiftparm(\rx) \right),
\label{eq:roc_gen}
\end{split}
\shortintertext{where}
	\eshiftparm(\rx) &= \frac{\mu_1(\rx) - \mu_0(\rx)}{\sigma_1(\rx)}, \nonumber \\ 
	\zeta(\rx) &= \frac{\sigma_0(\rx)}{\sigma_1(\rx)}. \nonumber
\end{align}
Thus, the ROC curve is completely determined by the shift and scale terms of the model. 

The binormal \citep{dorfman1969maximum} and bilogistic
\citep{ogilvie1968maximum} ROC curves can be obtained by setting~$\FZ$ to the standard normal distribution function~$\probit^{-1}=\Phi$, or the standard logistic distribution function~$\logit^{-1}(x) =\expit(x)=(1+\exp(-x))^{-1}$, in Equation~\ref{eq:roc_gen}, respectively. 
Similarly, the proportional hazard~\citep{gonen2010lehmann} and reverse proportional hazard alternatives~\citep{khan2022} for the ROC curve also fall within the purview of our transformation model with $\FZ$ specified as $\cloglog^{-1}(x)=1-\exp(-\exp(x))$ (minimum extreme value distribution function) and $\loglog^{-1}(x)=\exp(-\exp(-x))$ (maximum extreme value distribution function), respectively. 

However, to the best of our knowledge, the only literature where the transformation function~$h$ is included in the model formulation of the ROC curve is~\AYcite{Zou}{zou1997analysis}, who jointly models the shift term and the parameters of a Box-Cox power transformation function.
A key point of this paper is that we explicitly estimate $h$ jointly with $\mu(\rx)$ from the observed data and are not restricted to normality imposed by power transformation families. Thus, the methods we propose allow for proper propagation of uncertainty from the estimated transformation function $\hhat$ into the estimates of the shift and scale terms of the model.

The ROC curve in Equation~\ref{eq:roc_gen} follows a parametric model depending on $\FZ$, but is distribution-free as in \AYcite{Pepe et al.}{pepe2000interpretation, alonzo2002distribution}, because no assumptions are made about the transformation~$h$ and consequently for the distribution of the test results.
The approach to model the test results as a function of the disease status and
covariates was originally proposed in the latent variable ordinal regression
setting by~\AYcite{Tosteson et al.}{tosteson1988general} and extended by~\AYcite{Pepe}{pepe1997regression, pepe2000interpretation, pepe2003statistical} to modeling covariate effects directly on the ROC curve. 

\AYcite{Tosteson et al.}{tosteson1988general} pointed out that to ensure
concavity of the induced ROC curve, the scale term must be omitted, i.e.,
$\sigma_d(\rx)=1$ for $d=\{0,1\}$. The ROC curve is termed \emph{proper} if it is concave or, equivalently, if the derivative of the ROC curve is a monotonically decreasing function~\citep{egan1975signal, pan1997proper, dorfman1997proper}. A concave ROC curve is desirable as it yields the maximal sensitivity for a given value of specificity~\citep{mcintosh2002combining}. In this sense, as the decision criterion for classifying subjects is optimal when
the ROC curve is concave, we focus the remaining work on the model involving
only the shift term. Hence, the effect of covariates on the ROC curve is contained in the difference between the shift terms for diseased and nondiseased subjects, $\eshiftparm(\rx) = \mu_1(\rx) - \mu_0(\rx)$. For a relaxation of this assumption, see~\AYcite{Siegfried et al.}{siegfried2022} who estimate the scale functions through regression models.

\subsubsection{Two-sample case}
To fix ideas, we first consider the case of two samples without covariates. Let
the shift term take the form~$\mu_d(\rx)= \eshiftparm d$. The CDF of the test
results in the nondiseased population is given by~$F_0(y)=F_Z(h(y))$ and in the
diseased population by~$F_1(y)=F_Z(h(y) - \eshiftparm)$. Using
Equation~\ref{eq:roc_gen}, the induced ROC curve can be expressed as
\begin{align}
\ROC(p) 
&=1-\FZ \left( \FZinv(1-p) - \eshiftparm \right).
\label{eq:roc_simple}
\end{align}
The model assumption implies that a monotone function~$h$ exists to
transform both~$Y_1$ and~$Y_0$ into the same distribution,~$Z \sim \FZ$ separated by a shift parameter,~$\eshiftparm$. The induced ROC curve
from this model doesn't assume a particular distribution of the test result,
rather, it quantifies the \emph{difference} between the test result
distributions on the scale of a user-defined $\FZ$. In this sense, the difference
between the test result distributions is described by~$\eshiftparm$. Each choice of $\FZ$ leads to a different interpretation
of $\eshiftparm$. For example, when~$\FZ$ is selected to be the standard normal distribution function, $\eshiftparm$ is the difference in means of the transformed test results comparing the diseased and nondiseased groups, $\E[h(Y_1) - h(Y_0)]$. Similarly, when $\FZ$ is the standard logistic distribution function, $\exp(\eshiftparm)$ is the ratio of odds of having a positive test result comparing diseased and nondiseased groups.
Closed-form expressions can be derived for summary indices of the ROC curve by solving the appropriate integrals. The expressions of AUC, $J$, the optimal threshold $c^*$, sensitivity and specificity at $c^*$ are given
for some well known choices of~$\FZ$ in Table~\ref{tab:sum_simple}.

\begin{table} \centering
	\resizebox{\columnwidth}{!}{
		\begin{tabular}{ccccc} \toprule 
			\diagbox[innerwidth = 1.5cm, height = 1cm]{Index}{$\FZ$}& $\probit^{-1}$ & $\logit^{-1}$ & $\cloglog^{-1}$ & $\loglog^{-1}$ \\ \midrule \addlinespace
			$\AUC$ & $\Phi \left( \frac{\eshiftparm}{\sqrt{2}} \right)$ & $\left\{\begin{array}{ll}
				\frac{\exp(\eshiftparm)(\exp(\eshiftparm) - 1 - \eshiftparm)}{(\exp(\eshiftparm) - 1)^{2}} & \eshiftparm \neq 0 \\
				1 / 2 & \eshiftparm = 0
			\end{array} \right.$ & \multicolumn{2}{c}{$\expit(\eshiftparm)$} \\ \addlinespace[10pt]
			$J$ & $1- 2\Phi \left( \frac{-\absshift}{2} \right)$ & $1 - 2\expit \left( \frac{-\absshift}{2} \right)$ & \multicolumn{2}{c}{$\exp\left( \frac{-\absshift}{e^{\absshift}-1} \right) - \exp\left( \frac{\absshift}{e^{-\absshift}-1} \right)$} \\ \addlinespace[10pt]
			$c^*$ & \multicolumn{2}{c}{$\hinv\left( \frac{\eshiftparm}{2} \right)$} & $\hinv \left( \log \left( \frac{\eshiftparm}{1-e^{-\eshiftparm}} \right) \right)$ & $\hinv \left( \log \left( \frac{e^{
			\eshiftparm} - 1}{\eshiftparm} \right) \right)$\\ \addlinespace[10pt]
			$\Sens(c^*)$ & \multirow{ 2}{*}{$\Phi \left( \frac{\eshiftparm}{2} \right)$} & \multirow{ 2}{*}{$\expit \left( \frac{\eshiftparm}{2} \right)$} & $\exp\left( \frac{-\eshiftparm}{e^{\eshiftparm}-1} \right)$ & $1-\exp \left( \frac{\eshiftparm}{e^{-\eshiftparm}-1} \right)$\\ \addlinespace[10pt]
			$\Spec(c^*)$ & & & $1-\exp \left( \frac{\eshiftparm}{e^{-\eshiftparm}-1} \right)$ & $\exp \left( \frac{-\eshiftparm}{e^{\eshiftparm}-1} \right)$\\
			\addlinespace[10pt] \bottomrule
		\end{tabular}
	}
\caption{Closed-form expressions for the area under the receiver operating characteristic curve (AUC), Youden Index ($J$), optimal classification threshold ($c^*$), sensitivity ($\Sens$) and specificity ($\Spec$) at $c^*$ in terms of the shift parameter $\eshiftparm$ in the linear transformation model given by $F_d(y)= \FZ(h(y) - \eshiftparm d)$.}
\label{tab:sum_simple}
\end{table}

\subsubsection{Conditional ROC curve}
The accuracy of a diagnostic test may be influenced by a set of covariates~$\rX$. To evaluate this effect on the ROC curve and its summary indices, we assume a linear transformation model with a shift term that takes the form
\begin{align}
	\label{eq:lp_int}
	\mu_d(\rx) = \eshiftparm d +  \rx^\top \xparm + d \rx^\top \intparm,
\end{align}
where $\xparm, \intparm \in \RR^P$ are the coefficient vectors for the covariates and interaction term, respectively.
Under this model, the resulting covariate-specific ROC curve is
\begin{align*}
	\ROC(p \mid \rx) = 1-\FZ \left(\FZinv(1-p) - (\eshiftparm + \rx^\top \intparm) \right),
\end{align*}
where the covariate effect on the ROC curve is given by the difference in shift terms between diseased and nondiseased subjects, $\eshiftparm(\rx) = \eshiftparm + \rx^\top \intparm$. 
Similarly, the covariate-specific AUC is given by
\begin{align}
\label{eq:auc_x}
	 \AUC(\rx) = \Prob(Y_0 < Y_1 \mid \rX=\rx) = a \left( \eshiftparm(\rx) \right) = a\left( \eshiftparm + \rx^\top \intparm \right),
\end{align}
where  $a: \RR \mapsto [0,1]$ is the AUC function from the first row of Table~\ref{tab:sum_simple} for different choices of $\FZ$.
The expressions for~$J$, $c^*$, sensitivity and specificity can analogously be adjusted to account for covariates, with~$\eshiftparm$ replaced by $\eshiftparm + \rx^\top \intparm $ in Table~\ref{tab:sum_simple}.
Interpretation of the interaction coefficient is similar to~\AYcite{Pepe}{pepe1998three}.
For example, in the case of a continuous covariate $X=x \in \RR$, for each possible specificity value $1-p \in (0,1)$, a unit increase in~$x$ results in a~$\eintparm$-unit increase in the ROC curve (or an increase in the sensitivity) on the scale of~$\FZ$.
If $\eintparm$ is positive, an increase in~$x$ corresponds to an increase in the ROC curve, indicating that a test is better able to discriminate the two populations for larger values of $x$ and, vice versa. Note that the ROC curve varies with the covariate contingent upon the presence of an interaction between $d$ and~$x$~\citep{pepe2003statistical}. For~$\eintparm=0$, the covariate affects the distribution of the test results from the diseased and nondiseased population, but not the ROC curve. That is, for all levels of~$x$, the difference between the transformed distributions $h(Y_1)$~and~$h(Y_0)$ is given by~$\eshiftparm$, thus the ROC curve is unchanged.
Analogous interpretations hold when we are interested in modeling a set of covariates~$\rX$, which could possibly include categorical covariates.

Standard regression techniques have also been proposed as an alternative to assess the effect of covariates on summary indices rather than deriving the induced ROC curve. For example,~\AYcite{Dodd et al.}{dodd2003partial} model the partial AUC as a regression function of covariates. 
Our model equivalently results in a regression model for the AUC when the shift term is given from Equation~\ref{eq:lp_int} and a particularly chosen scale~$\FZ$.
Specifically, note that Equation~\ref{eq:auc_x} is the model proposed by~\AYcite{Dodd et al.}{dodd2003partial}, where~$\eshiftparm + \rx^\top \intparm$ is in the form of a usual linear predictor and $a$ is a monotonically increasing inverse link function which defines the scale for the regression coefficients.
This relationship provides another interpretation of the interaction term. 
For example, when $\FZ = \cloglog^{-1}$ or $\FZ = \loglog^{-1}$, $\exp(\eintparm)$ is a ratio of AUC odds comparing a subject with covariate value $x +1$ to $x$. The AUC odds conditional on~$x$ are defined as~${\Prob(Y_1 > Y_0 \mid X=x)}/{\Prob(Y_1 \leq Y_0 \mid X=x)}$. 
As will be shown in Section~\ref{sec:uniest}, an advantage of our approach is that we estimate the regression parameters of the transformation model using maximum likelihood estimation and do not rely on less efficient binary regression techniques. Thus, we retain parameter interpretation while allowing for efficient estimation of a variety of covariate-specific summary indices of the ROC curve.

We show that our method is additionally related to the probabilistic index model (PIM) of \AYcite{Thas et al.}{thas2012probabilistic, de2019semiparametric}, who propose to model the AUC as a function of $\rX^*-\rX$. Here, the superscript (*) indicates an alternate configuration of the modeled covariates. We discuss the PIM of the form 
\begin{align*}
	\Prob(Y < Y^* \mid D, D^*, X, X^*) = a\left( \eshiftparmtilde(D^*-D) + (\rX^* - \rX)^\top \xitildevec + (D^* \rX^* - D \rX)^\top \gammatildevec \right)
\end{align*}
where $a$ is the function as defined previously and the tilde ($\texttildelow$) is used to distinguish the parameters of the PIM from those of the transformation model.
Contrasting a diseased and nondiseased subject with the same covariate value, we let $D^*=1$, $D=0$, $\rX^* = \rX$ and $a=\logit^{-1}$, then the probabilistic index is given by  $\expit(\eshiftparmtilde + \rX^\top \gammatildevec)$. Thus, this is equivalent to the AUC resulting from a linear transformation model with $\FZ=\cloglog^{-1}$ and a shift term as in Equation~\ref{eq:lp_int}. Similarly, when $\FZ=\probit^{-1}$, we have the following relationship between the parameters of the linear transformation model and the PIM, $\eshiftparmtilde = \eshiftparm/ \sqrt{2}$ and $\gammatildevec = \intparm /\sqrt{2}$ . 
As was the case for ROC curves modeled by transformation models, the interaction between $D$ and $\rX$ allows the AUC to vary with covariates. The main effect $\xitildevec$, however, does not effect the AUC for configurations where the interest is to compare the distributions of $Y_1$ and $Y_0$ among subjects with a specific value of covariates.

We can also consider more general and potentially nonlinear formulations of the shift and scale terms in our framework. For the special case of $\FZ = \probit^{-1}$, the AUC from a shift-scale transformation model \citep{siegfried2022} is given by
\begin{align*}
	\Prob(Y_0 < Y_1 \mid \rX_0=\rx_0, \rX_1=\rx_1) = \Phi \left( \frac{\mu_1(\rx_1) - \
		\mu_0(\rx_0) }{\displaystyle \sqrt{\sigma_0(\rx_0)^2 + \sigma_1(\rx_1)^2}} \right)	
\end{align*}
where $\rX_0$ and $\rX_1$ are the corresponding (potentially different) sets of covariates in the nondiseased and diseased populations, respectively. 
When the scale term depends only on a single set of covariates,~$\sigma_0(\rx_0)=\sigma_1(\rx_1)=\sigma(\rx)$, despite varying sets of covariates in the shift terms, all the expressions hold from Table~\ref{tab:sum_simple} with $\eshiftparm$ replaced by~$\frac{\mu_1(\rx_1) - \mu_0(\rx_0)}{\sigma(\rx)}$.
However, such closed-form expressions cannot be derived for other choices of $\FZ$ when the scale term depends on the disease indicator or on different sets of covariates. In such cases, AUCs and other summary indices can be derived using numeric techniques on the induced ROC curve.

\subsubsection{Comparing correlated ROC curves}
If two (or more) diagnostic tests are measured on the same set of subjects (paired design), the test results are likely to be correlated leading to paired or correlated ROC curves \citep{swets2012evaluation}. To compare the accuracy of these tests, an approach that is frequently used in practice is to consider a summary index of the ROC curve and evaluate a hypothesis for the equivalence of the indices. For example, \AYcite{DeLong et al.}{delong1988comparing} developed an asymptotically normal nonparametric statistic to compare correlated AUCs based on the theory of U-statistics; see also~\AYcite{Hanley and others}{hanley1983method, wieand1989family, thompson1989statistical, mcclish1989analyzing, jiang1996receiver, bandos2005permutation, bandos2006permutation} who provide similar methods.
The problem with using the AUC as a summary index is that diagnostic tests with different ROC curves could potentially have the same value for the AUC.
To overcome this limitation, we construct a distribution-free test for differences in the marginal ROC curve parameters using multivariate transformation models that inherently account for the correlations between outcomes in their dependency structure.
The null hypothesis represents the equality of entire ROC curves instead of comparing correlated summary indices. Additionally, our test allows for covariate adjusted comparisons to be made.

Let the continuous random vector  $\mY = (Y_1$, $Y_2,\dots, Y_J)$ denote the results of $J$ diagnostic tests, possibly dependent on a set of covariates~$\rX$. This notation should not be confused with subscripts denoting quantities conditional on $D=1$.
To account for the correlation between the test results, we consider a multivariate transformation model with the strictly monotonically increasing conditional transformation function $g:\RR^J \mapsto \RR^J$. This function maps the unknown conditional distribution of $\mY$ depending on the disease indicator and covariates, to a random vector with a known distribution $\mZ = g(\mY \mid d, \rx)$.
Specifically, the vector $\mZ = (Z_1,\dots,Z_J )$ follows a zero mean multivariate normal distribution $\mZ \sim N_J(\nullvec, \mSigma)$ with covariance matrix $\mSigma \in \RR^{J \times J}$. 
Consequently, the marginal distribution of a multivariate normal vector is itself normally distributed with $Z_j \sim N(0, \varsigma^2_j)$, where the diagonal elements of $\mSigma$ give the variances~$\varsigma_j^2$ for $j=1,\dots,J$.
Thus, the conditional joint CDF of $\mY$ is given by
\begin{align*}
	\Prob(\mY \leq \yvec \mid D=d, \rX=\rx) &= \Prob(g(\mY \mid d, \rx) \leq g(\yvec  \mid d, \rx) ) \\
	&= \Prob(\mZ \leq g(\yvec  \mid d, \rx) ) \\
	&= \mathlarger{\Phi}_{\nullvec, \mSigma} \left(g_1(y_1 \mid d, \rx ), \dots, g_J(y_J \mid d, \rx)  \right)
\end{align*}
where {\large $\Phi_{\nullvec, \mSigma}$} is the joint CDF of a multivariate normal distribution with a zero mean vector and covariance matrix $\mSigma$. Letting
\begin{align*}
\label{eq:gj}
	g_j(y_j \mid d, \rx ) &= \Phi^{-1}_{0, \varsigma_j^2} \left( F_Z(h_j(y_j) - \mu_d(\rx)) \right)
\end{align*}
where $\Phi_{0, \varsigma^2}$ is the CDF of a normal distribution with variance $\varsigma^2$, the CDF of $\mY$ coincides to the Gaussian copula and the marginal distribution of $Y_j$ is given by
\begin{align*}
	\Prob(Y_j \leq y_j  \mid D=d, \rX = \rx) &= \mathlarger{\Phi}_{\nullvec, \mSigma} \left(\infty, \dots, \infty, g_j(y_j \mid d, \rx ), \infty, \dots, \infty \right) \\
	&= 	\Phi_{0, \varsigma_j^2} \left( 	g_j(y_j \mid d, \rx ) \right) \\
	&= \FZ \left( h_j(y_j) - \mu_{d,j}(\rx) \right).
\end{align*}
By using the multivariate model specified we allow $\mY$ to have a nonlinear dependence structure through the marginal transformation functions $h_j$. Analogous to our univariate transformation model, this multivariate model implies that the~$j$th marginal transformed test results follow
\begin{align*}
	h_j(Y_j) = \mu_{d,j}(\rx) + Z
\end{align*}
where the function $\mu_{d,j}$ is the shift term for the $j$th test result model and $Z \sim \FZ$, as described previously.
The marginal ROC curve is then given by
\begin{align*}
	\ROC_j(p \mid \rx) = 1 - \FZ \left( \FZinv(1-p) - \eshiftparm_j(\rx) \right)
\end{align*}
where $\eshiftparm_j(\rx)=\mu_{1,j}(\rx) - \mu_{0,j}(\rx)$ and values of marginal summary indices can be computed from the closed-form expressions given in Table~\ref{tab:sum_simple} using the marginal ROC curve parameters. Specifically, $\delta$ can be replaced by the covariate effect on the summary indices given by $\eshiftparm_j(\rx)$. From this result, we see that the null hypothesis of equal ROC curves at a given set of covariate values $\rx$ for the $j$th and $k$th test is equivalent to testing if $\eshiftparm_j(\rx)-\eshiftparm_k(\rx)=0$. Moreover, this hypothesis test is distribution-free as we make no assumptions about $h_j$.

\AYcite{Klein et al.}{klein2022multivariate} establish conditions which decompose general transformation functions into the marginal ones described above. Specifically, the covariance matrix takes the form $\mSigma = \mLambda^{-1} \mLambda^{-\top}$ where $\mLambda$ is a lower triangular $(J \times J)$ matrix whose coefficients characterize the dependence structure given by the Gaussian copula. The matrix~$\mLambda$ could also vary between diseased and nondiseased subjects and potentially depend on a set of covariates $\rx$. This allows the dependence structure of $\mY$ to change as a function of $\rx$. For the special case of two diagnostic tests, the correlation between $Y_1$ and $Y_2$ for a given set of covariates $\rx$ is given by $\frac{-\lambda_d(\rx)}{\sqrt{\lambda_d(\rx)^2 + 1}}$. For more details and examples of multivariate transformation models, see~\AYcite{Klein and others}{klein2022multivariate, Barbanti2022}.

\subsection{Univariate estimation}
\label{sec:uniest}
In this section we propose estimation methods for a transformation model with univariate test results. We provide an explicit parameterization of the transformation function and the shift term. We then maximize the likelihood contributions for a potentially exact continuous, right-, left- or interval-censored datum to jointly estimate the model parameters.
This enables us to fully determine the ROC curve and its summary indices as well as handle test results which are ordinal or impacted by instrument detection limits.

\subsubsection{Parameterization}
We parameterize the transformation function as
\begin{align}
	h(y \mid \parm) = \basisy(y)^\top \parm = \sum_{m=0}^{M} \vartheta_m b_m(y) \quad \text{for} \quad y \in \RR, 
\end{align}
where~$\basisy(\ry)=(b_0(\ry),\dots,b_M(\ry))^\top$ is vector of $M+1$ basis
functions with coefficients~$\parm \in \RR^{M+1}$.
Polynomials in Bernstein form offer a choice of basis that provides a flexible way of estimating the underlying function. The Bernstein basis polynomial of order~$M$ is defined on the
interval~$[l,u]$ as
\begin{align}
\label{eq:basis}
	b_m(y) = {M \choose m} \tilde{y}^m (1-\tilde{y})^{M-m}, \quad m=0,\dots,M,
\end{align}
where $ \tilde{y} = \frac{y-l}{u-y} \in [0,1]$. The restriction $\vartheta_m
\leq \vartheta_{m+1}$ for $m=0,\dots,M-1$, guarantees the monotonicity of $h$.
Observe that the transformation function is linear in the parameters that define
it and, any nonlinearity of the test results is modeled by the basis functions.
If the order $M$ is chosen to be sufficiently large, Bernstein polynomials can
uniformly approximate any real-valued continuous function on an interval \citep{farouki2012bernstein}.

\subsubsection{Likelihood}
\label{sec:uni_likelihood}

Denote the complete parameter vector as $\thetavec=(\shiftparm^\top, \parm^\top)^\top$,
where $\shiftparm=(\eshiftparm, \xparm^\top, \intparm^\top)^\top \in \RR^{2P+1}$ are the vector of regression coefficients parameterizing the function $\mu_d$ from Section~\ref{sec:trafo} and $\parm \in \RR^{M+1}$ are the basis coefficients. We follow the maximum likelihood approach proposed by \AYcite{Hothorn et al.}{hothorn2018most} to
jointly estimate~$\shiftparm$~and~$\parm$. The advantages of embedding the model
in the likelihood framework are as follows. (i)~All forms of random censoring
(right, left, interval) as well as truncation can directly be incorporated into
likelihood contributions defined in terms of the distribution function
\citep{lindsey1996parametric}. (ii)~If the given model is correctly specified,
under regularity conditions, the maximum likelihood estimator (MLE) will be
asymptotically the most efficient estimator (minimum variance and unbiased).
(iii)~The MLE is asymptotically normally distributed and has a sample variance
that can be computed from the inverse of the Fisher information matrix. This can
be used to generate confidence intervals for the estimated parameters. (iv)~The
MLE is equivariant which implies invariance of the score test (or the Lagrange
multiplier test) to reparameterizations
\citep{rao1948large,aitchison1958maximum, silvey1959lagrangian,
dagenais1991invariance}. Specifically, as will be shown in
Section~\ref{sec:score}, by inverting the score test, our method produces
confidence bands for the ROC curve and appropriate score intervals for its
summary indices.

The likelihood contribution of a single observation~$O=(Y,D,\rX)$ where $Y \in
(\underline{y},\overline{y}]=\{ y\in \RR : \underline{y} < y \leq
\overline{y}\}$ is given by
\begin{equation*}
L(\thetavec \mid O) = 
\begin{cases}
\begin{aligned}
&f_Z(h(y \mid \parm) - \mu_d(\rx \mid \shiftparm)) h'(y \mid \parm) && y \in \RR &\quad&  \text{`exact continuous'} \\
&1-\FZ(h(\underline{y} \mid \parm) - \mu_d(\rx \mid \shiftparm)) && y \in (\underline{y}, \infty) &\quad& \text{`right censored'} \\
&\FZ(h(\overline{y} \mid \parm) - \mu_d(\rx \mid \shiftparm)) && y \in (-\infty, \overline{y}) &\quad& \text{`left censored'} \\
& \begin{split}
	\FZ(h(\overline{y} \mid \parm) - \mu_d(\rx \mid \shiftparm)) - \\
	\FZ(h(\underline{y} \mid \parm) - \mu_d(\rx \mid \shiftparm))
\end{split}  && y \in (\underline{y}, \overline{y}] &\quad& \text{`interval censored'}, \\
\end{aligned}
\end{cases}
\end{equation*}
where $f_Z$ is the density function of $Z$ and $h'(y \mid \parm)$ is the first
derivative of the transformation function with respect to $y$. Given a sample of
$N$ independent and identically distributed observations
$O_i$~for~$i=1,\dots,N$, the log-likelihood is given by
$\ell(\thetavec)=\sum_{i=1}^N \log(L_i(\thetavec))$, where $L_i$ is the likelihood
contribution of observation $i$. The (unconditional) maximum likelihood estimate
of~$\thetavec$ is the solution to the optimization problem
\begin{align*}
	\hat{\thetavec}=(\hshiftparm, \hparm) = \argmax_{\shiftparm, \parm}  \ell(\shiftparm, \parm), 
\end{align*}
subject to the monotonicity constraint $\vartheta_m \leq \vartheta_{m+1}$ for
$m=0,\dots,M-1$. The resulting ROC curve only depends on $\shiftparm$ which is decoupled from the parameters needed to model the transformation function~$\parm$. The score function is defined as the first derivative of the log-likelihood function with respect to each of the parameters and is given by
\begin{align*}
	S(\thetavec) = 
	\left( \begin{array}{l}
		\frac{\partial \ell(\thetavec)}{\partial \shiftparm}\\
		\frac{\partial \ell(\thetavec)}{\partial \parm}
	\end{array} \right) =
	\left( \begin{array}{l}
		S_\shiftparm(\thetavec) \\
		S_\parm(\thetavec)
	\end{array} \right).
\end{align*}
We perform constrained optimization using the likelihood and score contributions to determine the maximum likelihood estimates for $\shiftparm$ and $\parm$. For computational details, see~\AYcite{Hothorn}{hothorn2020most}.
 The asymptotic variance of the MLE can further be estimated by the expected Fisher information matrix which is the variance-covariance matrix of the score function and is defined as
\begin{align*}
	I(\thetavec) = - \E
	\left(\begin{array}{ll}
		\frac{\partial^2 \ell(\thetavec)}{\partial \shiftparm \partial \shiftparm^\top} &
		\frac{\partial^2 \ell(\thetavec)}{\partial \shiftparm \partial \parm^\top} \\
		\frac{\partial^2 \ell(\thetavec)}{\partial \parm \partial \shiftparm^\top} &
		\frac{\partial^2 \ell(\thetavec)}{\partial \parm \partial \parm^\top}
	\end{array} \right) = 
	\left(\begin{array}{ll}
		I_{\shiftparm, \shiftparm}(\thetavec) & I_{\shiftparm, \parm}(\thetavec) \\
		I_{\shiftparm, \parm}(\thetavec)^\top &
		I_{\parm, \parm}(\thetavec)
	\end{array} \right).
\end{align*}
The matrix is partitioned such that the submatrix $I_{\shiftparm, \shiftparm}(\thetavec)$ corresponds to the parameter related to the disease indicator and covariates. The variation in~$\parm$ and its relationship to~$\shiftparm$ is not of direct relevance here but is nonetheless estimated.

\subsubsection{Limit of detection}
Instrument precision can affect the evaluation of diagnostic biomarkers. For example, when biomarker levels are at or below the limit of detection (LOD) $y_{\LOD}$, the observed value lies in an interval~$(-\infty, y_{\LOD})$ and the resulting measurement is left censored. Often a replacement value is substituted for such measurements. Alternatively, only biomarker values above the LOD are used for the ROC analysis. It has been shown that these approaches lead to biased estimation \citep{lynn2001maximum, singh2002robust}. Various adjustments to ROC curves and its summary indices have been proposed to handle such censored measurements~\citep{mumford2006pooling, perkins2007receiver, perkins2009generalized, ruopp2008youden, bantis2017estimation, xiong2022family}. However, these methods typically do not account for covariates.
Our framework naturally accounts for such observations in the likelihood function for left censored test results. Similarly, the right censored likelihood accounts for measurements which are affected by an upper limit of detection.
Thus, our method provides a smooth covariate-specific ROC curve for all values of specificity with estimates and inference appropriately incorporating the observed information.

\subsubsection{Ordinal test results}
Many diagnostic tests of interest to clinical researchers are measured on ordinal scales. These are rank-ordered categorical variables for which quantitative differences between levels are unknown. 
Examples include the Memory Impairment Screen (MIS) for the measurement of cognition to distinguish Alzheimer's disease and other dementias \citep{buschke1999screening} or the risk assessment of disease from radiological images such as with the Breast Imaging Reporting and Data System (BI-RADS) \citep{birads2013}. 
A simple reparameterization of the transformation function facilitates an extension of ROC curves to such ordinal test results.

Let the random variable $Y \in \{y_1,\dots,y_K\}$ denote the results of an ordered categorical diagnostic test with $K$~categories, where $y_k < y_{k+1}$ for $k=1,...,K-1$ . Define the covariate-specific ROC curve as a discrete function of the subject categories
\begin{align*}
	\left\lbrace \left(1-F_0(y \mid \rx), 1-F_1(y \mid \rx) \right), y\in \{y_1,\dots,y_K\} \right\rbrace.
\end{align*}
This set of points plots the covariate-specific $1-\text{specificity}$ and sensitivity coordinates for each of the $K$~categories. Suppose that the following transformation model defines the distribution of the ordinal test results in the diseased and nondiseased populations
\begin{align*}
	F_d(y_k \mid \rx) = F_Z \left( h(y_k) - \mu_d(\rx) \right),
\end{align*}
where the transformation function assigns one parameter to each of the categories of the test result except the last one, $h(y_k) = \vartheta_k \in \RR$ for $k=1,\dots,K-1$. Note that since $\Prob(Y \leq y_K \mid \rX = \rx) = 1$, only~$K-1$ parameters need to be specified. The monotonicity constraint $\vartheta_k < \vartheta_{k+1}$ is also necessary so that $F_d(y_k \mid \rx) < F_d(y_{k+1} \mid \rx)$ for $k=1,\dots,K-1$. Observe that this is a special case of the general shift-scale model in Equation~\ref{eq:cdf_gen}. Under a linear shift term as given in Equation~\ref{eq:lp_int} and $\FZ = \logit^{-1}$, this model is known as the proportional odds logistic regression (POLR). 

Setting $p_k = 1-F_0(y_k \mid \rx)$, the ROC curve under the ordinal transformation model is given by
\begin{align*}
	\ROC(p_k) = 1-\FZ \left( \FZinv(1-p_k) - \eshiftparm(\rx) \right), \quad k=1,\dots,K.
\end{align*}
The parameters of the model can be estimated by using the maximum likelihood scheme described previously with the interval censored likelihood contribution
\begin{align*}
	\FZ(\vartheta_k - \mu_d(\rx \mid \shiftparm) ) - \FZ(\vartheta_{k-1} - \mu_d(\rx \mid \shiftparm)),
\end{align*}
where the observed subject has a test result of $y_k$ and covariate value $\rx$. After estimating the parameters, it is possible to plot a continuous ROC curve as a function of all $p\in [0,1]$. This curve's shape depends on the choice of $\FZ$ and represents the underlying parametric form of the model. However, it does not represent the ROC curve for a continuous test. Specifically, the function is only defined at the $K$ discrete specificity values. Summary measures can be calculated using the expressions detailed in Table~\ref{tab:sum_simple}. Note that these would be based on a smooth interpolation given by the model rather a discrete one and lead to the interpretation of an ordinal test with a latent continuous scale.

\subsection{Multivariate estimation}
The estimation of a multivariate test result follows a similar approach to the univariate case. The marginal transformation functions are parameterized in terms of monotonically increasing basis polynomials of Bernstein form $h_j(y_j \mid \parm_j) = \basisy_j(y_j)^\top \parm_j $, where $\basisy_j : \RR \mapsto \RR^{M+1}$ is a vector of $M+1$ basis functions with individual components $b_{j,m}$ defined as in Equation~\ref{eq:basis} for $j=1,\dots,J$ and $m=0,\dots,M$. The constraint $\vartheta_{j,m}
\leq \vartheta_{j,m+1}$ for~$m=0,\dots,M-1$, guarantees the monotonicity of $h_j$ and the smooth parameterization the existence of the derivative $\h'_j(y_j \mid \parm_j) = \basisy'_j(y_j)^\top \parm_j$.

With a slight abuse of notation compared to the univariate case, we also denote the vector of regression coefficients as $\shiftparm=(\shiftparm_1^\top,\dots,\shiftparm_J^\top)^\top \in \RR^{J(2P+1)}$
and the vector of basis coefficients as $\parm=(\parm_1^\top,\dots,\parm_J^\top)^\top \in \RR^{J(M+1)}$, with each component parameterizing a separate marginal conditional distribution function. The covariance matrix $\mSigma$ captures the correlation structure between the transformed diagnostic tests and the vector $\varsigmavec \in \RR^{J(J+1)/2}$ contains all the elements of~$\mSigma$. We denote the complete set of unknown model parameters for a transformation model with a multivariate test result as $\thetavec = (\shiftparm^\top, \parm^\top, \varsigmavec^\top )^\top$.
The likelihood contribution of a single observation $O=(\mY, D, \rX)$, with $\mY = \yvec = (y_1,\dots,y_J)^\top$ being an exact continuous vector of test results, is given by
\begin{align*}
	 L(\thetavec \mid O) = \mathlarger{\phi}_{\nullvec, \mSigma} \left( g(\yvec \mid d, \rx, \shiftparm, \parm) \right) \prod_{j=1}^J g'_j \left( y_j \mid d, \rx, \shiftparm, \parm \right),
\end{align*}
where {\large $\phi_{\nullvec, \mSigma}$} is the joint density function of a multivariate normal distribution with a zero mean vector and covariance matrix $\mSigma$. With a choice of each component $g_j$ from Equation~\ref{eq:gj}, its derivative is given by
\begin{align*}
	g'_j(y_j \mid d, \rx, \shiftparm, \parm) = \frac{f_Z \left( h_j(y_j \mid \parm) - \mu_d(\rx \mid \shiftparm) \right) h'_j(y_j \mid \shiftparm)}{\phi_{0,\varsigma_j^2} \left( \Phi^{-1}_{0,\varsigma_j^2} \left\{ \FZ \left( h_j(y_j \mid \parm) - \mu_d(\rx \mid \shiftparm) \right) \right\} \right)}.
\end{align*}
In the special case where $\FZ=\probit^{-1}$, this derivative simplifies to $\varsigma_j h'_j(y_j)$. 

We derive the maximum likelihood estimate of $\thetavec$ from a sample of $N$ independent and identically distributed observations using constrained maximization algorithms following the strategy detailed in Section~\ref{sec:uni_likelihood}. For more details see \AYcite{Klein et al.}{klein2022multivariate}.

\subsection{Confidence intervals}
In the following section, we present three methods to calculate confidence bands for the ROC curve and confidence intervals for its summary indices. These quantities are functions of the parameters in the model $G(\shiftparm): \RR^{2P+1} \rightarrow \RR$ and to maintain nominal coverage for a confidence interval for $G(\shiftparm)$, appropriate methods are needed.
The methods discussed include inversion of the score test, delta method and simulation from the asymptotic distribution of the estimate. The methods are ordered by their degree of theoretical justification. We start with score intervals which are invariant to parameter transformations but become computationally expensive when dealing with a large set of parameters. We then discuss estimating the variance using the multivariate delta method and conclude with a simple simulation method which is versatile without being computationally demanding.

\subsubsection{Score intervals}
\label{sec:score}

In the two-sample univariate case where $\eshiftparm$ defines the ROC curve, as in Equation~\ref{eq:roc_simple}, we can construct score intervals for $\eshiftparm$. Unlike the Wald and other commonly used intervals, score intervals are especially desirable as they are invariant to transformations of the parameters. That is, a score CI for $G(\eshiftparm)$ (e.g., the AUC $a(\eshiftparm)$), provides the same level of coverage as would a score CI for $\eshiftparm$.  In turn, under a correctly specified model, a score CI for $\eshiftparm$ allows the construction of accurately covered uniform confidence bands for the ROC curve as well as intervals for its summary indices such as the AUC and the Youden index.

We first generate score confidence intervals for~$\eshiftparm$ by inverting the score test. In this case, the null hypothesis is given by~$H_0:\eshiftparm=\eshiftparm_0$ where $\eshiftparm_0$ is a specific value of the parameter of interest. Under $H_0$, the restricted (conditional)
maximum likelihood estimator for $\parm$ can be obtained by
\begin{align*}
	\hparm(\eshiftparm_0)= \argmax_{\parm} \ell(\eshiftparm_0, \parm),
\end{align*}
or as a solution of the $M+1$ score equations
$S_\parm(\eshiftparm_0,\parm)=\nullvec$. Note that this estimate is a function of $\eshiftparm_0$.
Letting~$\thetatildevec=(\eshiftparm_0, \hparm(\eshiftparm_0))$, the quadratic
(Rao) score statistic simplifies to
\begin{align*}
	R(\eshiftparm_0) &= S(\thetatildevec)^\top I^{-1}(\thetatildevec) S(\thetatildevec) \\
	&= ( S_\eshiftparm(\thetatildevec)^\top, \nullvec^\top ) I^{-1}(\thetatildevec) ( S_\eshiftparm(\thetatildevec)^\top, \nullvec^\top )^\top \\
	&= S_\eshiftparm(\thetatildevec)^\top A_{\eshiftparm,\eshiftparm}(\thetatildevec) S_\eshiftparm(\thetatildevec),
\end{align*}
where $A_{\eshiftparm,\eshiftparm}(\thetatildevec)$ denotes the submatrix
corresponding to $\eshiftparm$ of the inverse Fisher information matrix and is given by the Schur complement $I_{\eshiftparm,\eshiftparm}(\thetavec) - I_{\eshiftparm,\parm}(\thetavec) I^{-1}_{\parm,\parm}(\thetavec) I_{\eshiftparm,\parm}(\thetavec)^\top$.
Under $H_0$, $R(\eshiftparm_0)$ converges asymptotically to a chi-square distribution
with 1 degree of freedom, $R(\eshiftparm_0) \Darrow \chi^2_1$. This result is
explained by \AYcite{Rao}{rao2005}.
Thus, inverting the score statistic by enumerating values of $\eshiftparm_0$
allows for the construction of~$(1-\alpha)$ score confidence intervals
for~$\eshiftparm$ defined as
\begin{align*}
	\{ \eshiftparm_0 \in \RR \mid R(\eshiftparm_0) < \chi^2_1(1-\alpha) \},
\end{align*}
where $\chi^2_P(1-\alpha)$ is the~$(1-\alpha)$ quantile value of the $\chi^2_1$
distribution. Equivalently, we can use the square root of the score statistic to
construct a $(1-\alpha)$ score interval using quantiles of the standard normal
distribution, $\{ \eshiftparm_0 \in \RR \mid \Phi^{-1}(\alpha/2)
<\sqrt{R(\eshiftparm_0)} \leq \Phi^{-1}(1-\alpha/2) \}$. Finally, we apply the function $G$ to both the lower and upper limits of the interval to construct score confidence bands for the ROC curve or score confidence intervals for its summary indices

The score statistic is given by~$R(\eshiftparm_0)=S_\eshiftparm(\thetatildevec)^2
A_{\eshiftparm,\eshiftparm}(\thetatildevec) $. Testing if there is a
significant difference between the nondiseased and diseased populations coincides to the hypothesis test, $H_0: \eshiftparm=0$. This is computationally
efficient because only the distribution of $R(0)$ needs to be computed. 
However, computing score CIs requires updating the restricted MLEs $\hparm(\eshiftparm_0)$ for an enumeration of $\eshiftparm_0$ values. This becomes computationally intractable when enumerating a higher dimensional grid of parameters.

\subsubsection{Delta method}
Since the MLE satisfies
\begin{align*}
	\sqrt{n} (\hshiftparm - \shiftparm) \xrightarrow{D} N_{P+1} \left( \nullvec, A_{\shiftparm,\shiftparm}(\thetavec) \right),
\end{align*}
then by the multivariate delta method, $G(\hshiftparm)$ also follows a normal distribution with
\begin{align*}
	\V(G(\hshiftparm)) = \frac{1}{n} \nabla G(\shiftparm)^\top A_{\shiftparm,\shiftparm}(\thetavec) \nabla G(\shiftparm),
\end{align*}
where $\nabla G(\shiftparm)$ is the gradient of $G$ evaluated at $\shiftparm$ and the inverse Fisher information matrix $A_{\shiftparm,\shiftparm}(\thetavec)$ is given by the Schur complement $I_{\shiftparm,\shiftparm}(\thetavec) - I_{\shiftparm,\parm}(\thetavec) I^{-1}_{\parm,\parm}(\thetavec) I_{\shiftparm,\parm}(\thetavec)^\top$. For example, when the shift term takes the linear form as in Equation~\ref{eq:lp_int} and $G$ defines the AUC function for $\FZ=\probit^{-1}$, the entries of $\nabla G(\shiftparm)$ are given by
\begin{align*}
	\frac{\partial  G(\shiftparm)}{\partial \eshiftparm} = \frac{1}{\sqrt{2}} C, \quad \frac{\partial  G(\shiftparm)}{\partial \exparm_i} = 0 \quad \text{and} \quad \frac{\partial  G(\shiftparm)}{\partial \eintparm_i} = \frac{x_i}{\sqrt{2}} C,
\end{align*}
where $C=\phi \left(\frac{\eshiftparm + \rx^\top \gammavec}{\sqrt{2}} \right)$, $\phi$ is the density of the standard normal distribution and $i$ indexes the~$P$ covariates. In general, the gradient can be estimated by calculating such derivatives and evaluating the resulting function at the MLE. Similarly, the variance-covariance matrix of the estimated parameters $A_{\shiftparm,\shiftparm}(\thetavec)$ can be computed by inverting the numerically evaluated Hessian matrix. Thus, a $(1-\alpha)$ level confidence interval for $G(\shiftparm)$ is given by
\begin{align*}
	G(\shiftparm) \pm \Phi^{-1}(\alpha/2) \sqrt{\hat{\V}(G(\hshiftparm))}.
\end{align*}

\subsubsection{Simulated intervals}
\label{sec:simints}
When the function $G$ has complex derivatives, as would be the case for nonlinear shift terms~$\mu_d(\rx)$ or when calculating optimal thresholds $c^*$ where $G$ includes the inverse of the transformation function, constructing confidence intervals using the delta method becomes infeasible. For these cases, we apply a simple simulation-based algorithm which utilizes the asymptotic normality of the MLE to calculate confidence intervals for the ROC curve and its summary indices, which are functions of the parameters of interest. The steps of the algorithm to construct $(1-\alpha)$ level confidence intervals for $G(\hshiftparm)$ can be summarized as follows:
\begin{enumerate}
	\item Generate $B$ independent samples from the asymptotic multivariate normal distribution of the parameter estimates $N_{P+1} \left( \hshiftparm, \frac{1}{n} \hat{A}_{\shiftparm,\shiftparm}(\mhatTheta) \right)$ and denote as $\hshiftparm^*_1,\dots,\hshiftparm^*_B$.
	\item For each sample $b=1,\dots,B$, calculate the function of interest $G(\hshiftparm^*_b)$.
	\item Construct the confidence interval $\left( Q_{G(\hshiftparm^*)}(\alpha/2), Q_{G(\hshiftparm^*)}(1-\alpha/2) \right)$, where $Q_{G(\hshiftparm^*)}$ is the empirical quantile function of the sample $G(\hshiftparm^*_1),\dots,G(\hshiftparm^*_B)$.
\end{enumerate}
A similar algorithm is presented in~\AYcite{Mandel}{mandel2013simulation}, who discuss its asymptotic validity and present several examples that show its empirical coverage adheres to nominal levels with results similar to the delta method.

\section{Empirical evaluation}
\begin{figure}[t!]
	\centering
	\includegraphics[width=\linewidth]{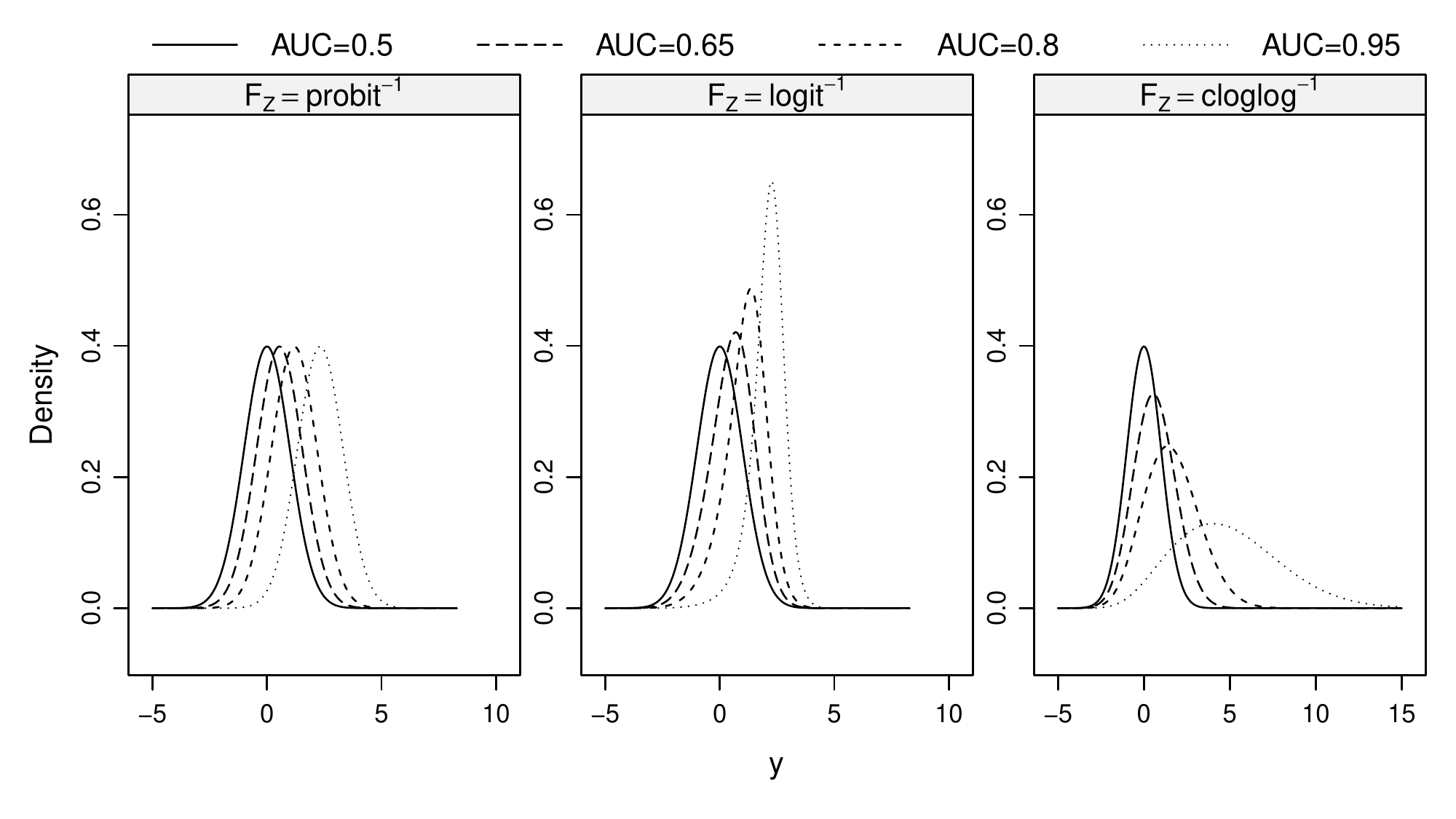} 
	\caption{Density functions for the model used to generate the data for the simulations. The nondiseased test results followed a standard normal distribution corresponding to an $\AUC=0.5$. The diseased test results varied with three choices of $\FZ$: $\probit^{-1}$, $\logit^{-1}$ and $\cloglog^{-1}$ each of which had an AUC of 0.5, 0.65, 0.8 and 0.95.}
	\label{fig:sim_dens}
\end{figure}

We conducted a simulation study to evaluate the performance of our estimators in the two-sample setting.  We considered a data generating process such that nondiseased test results followed a standard normal distribution $F_0(y) = \Phi(y)$ and the diseased results a distribution with the CDF $F_1(y) = \FZ( \FZinv(\Phi(y)) - \eshiftparm)$.
To obtain different shapes of the ROC curve, we chose three choices of $\FZ \in \{\probit^{-1}, \logit^{-1}, \cloglog^{-1} \}$ and varied~$\eshiftparm$ such that the $\AUC \in \{ 0.5, 0.65, 0.8, 0.95 \}$ or that $J \in \{ 0, 0.25, 0.5, 0.8 \}$, leading to a variety of configurations. 
Under this simulation paramaterization, the ROC curves followed the form of Equation~\ref{eq:roc_simple} and summary indices could be calculated as a function of the estimated $\eshiftparm$ from Table~\ref{tab:sum_simple}.
The conventional binormal model corresponded to $\FZ=\probit^{-1}=\Phi$, which induced proper binormal ROC curves. This was the only configuration where the test results for both groups were normally distributed. We included this configuration to ascertain the loss of power associated with our estimators when the standard binormal assumption held.
For other choices of $\FZ$ with $\AUC > 0.5$, the resulting distributions of the diseased test results were non-normal, with variances and higher moments differing between the two groups.
Specifically, the configuration of $\FZ=\logit^{-1}$ led to light tailed distributions for the diseased test results, whilst $\FZ=\cloglog^{-1}$ led to skewed, heavy-tailed distributions.
The corresponding density functions for the data generating model with selected AUC values are given in Figure~\ref{fig:sim_dens}.

For \numprint{10000} replications of each configuration, we generated balanced data sets with sample sizes $N_0=N_1 \in \{25, 50, 100\}$. The transformation models discussed in Section~\ref{sec:methods} were fitted to the simulated data sets assuming a parameterization of the transformation function given by a Bernstein basis polynomial of order $M=6$. The true data-generating model had a nonlinear transformation function $h=\FZinv \circ \Phi$. Our model estimation procedure aimed to approximate this function alongside the shift parameter $\eshiftparm$.
The functions implementing these models are available from the \pkg{tram} package \citep{hothorn2020most}, and included \code{BoxCox}, \code{Colr} and \code{Coxph} models which corresponded to $\FZ$ selected as $\probit^{-1}$, $\logit^{-1}$ and  $\cloglog^{-1}$, respectively.

Table~\ref{tab:rmse} displays the bias and root-mean-square error (RMSE) for the AUC estimates using different methods under the various configurations. We found that all three methods had minimal bias and equivalent RMSEs for an $\AUC=0.5$, where the test results were unable to distinguish between the two groups. The \code{BoxCox} and \code{Colr} models yielded approximately unbiased AUC estimates in all cases, even when they were misspecified for the true data generating process. However, estimates based on the proportional hazards model, \code{Coxph}, were biased for data generating processes other than~$\FZ=\cloglog^{-1}$.
We also computed nominal 90\% score confidence intervals for the AUC using the approach described in Section~\ref{sec:score}. The observed coverage probabilities are summarized in Table~\ref{tab:cvg}. When $\FZ$ was correctly specified, all methods had close to nominal coverage under all sample size configurations. In addition, the score confidence intervals from the \code{Colr} model were accurate even when it was misspecified for the true data generating process. This was unlike the \code{BoxCox} and \code{Coxph} models, whose coverage deteriorated for higher AUC values when they were misspecified.

We compared our approaches to a set of alternative methods for computing confidence intervals for the AUC and Youden index. The software details of all the methods used alongside their respective features and references are summarized in Supplementary Material Table~\SMref{S1}{tab:comp_mthds}. We detail the empirical coverage and average width of the confidence intervals for the AUC in Figures~\SMref{S1}{fig:cvg_auc} and~\SMref{S2}{fig:pwr_auc}, respectively. Estimates based on transformation models (\proglang{R} packages \pkg{tram}, \pkg{orm}, \pkg{pim}) yielded coverage close to the nominal level and significantly outperformed the other methods when the model was correctly specified for the true data generating process. All other methods generally performed close to nominal levels for low to medium AUC values (0.5-0.8), but broke down for higher AUC values. Methods which used $\FZ=\cloglog^{-1}$ gave confidence intervals which were shorter in length (overconfident) and otherwise there were no notable differences between the methods.

Analogously, Figures~\SMref{S3}{fig:cvg_yi} and~\SMref{S4}{fig:pwr_yi} detail the coverage and length of the confidence intervals for the Youden index. The methods which were based on the overlap coefficient failed to cover the configuration where $J=0$ because their lower limits were never below 0. Our methods estimated confidence intervals for $\eshiftparm \in \RR$ which naturally accounted for this scenario. Specifically, the lower limit of the CI for $J$ is given by $G(\max \{0, \eshiftparm^l \})$ and the upper limit by $G(\max \{ \lvert \eshiftparm^l \rvert, \lvert \eshiftparm^u \rvert \})$, where $G$ is the Youden index function defined in Table~\ref{tab:sum_simple} and $(\delta^l,\delta^u)$ is the confidence interval for $\delta$. Remarkably, the transformation model with~$\FZ=\logit^{-1}$ provided coverage at nominal levels for all simulation configurations with a relatively small confidence interval width. The approach of \cite{franco2021inference} (FP) was also accurate under model misspecification but was more involved. Namely, it consisted of estimating Box Cox transformation parameters under a binormal framework with bootstrap variance, all carried out on the logit scale and then back-transformed. In a setting with covariates, censoring or with $J=0$ this methodology would be limited.

Figures~\SMref{S5}{fig:cvg_roc} and~\SMref{S6}{fig:pwr_roc} show the coverage and area of the confidence bands for the ROC curve. All the approaches based on transformation models covered the configuration with $\AUC=0.5$ accurately. However, the other approaches did not yield coverage close to nominal levels in this configuration with the exception of \cite{martinez2018efficient}, whose confidence bands had a significantly larger area. For all other configurations, only transformation models which were correctly specified for the true data-generating model provided accurate results.

\begin{table}\centering
\begin{tabular}{llllcccccccc}
\toprule
&&&$\FZ$ & \multicolumn{2}{c}{$\text{probit}^{-1}$}&&\multicolumn{2}{c}{$\text{logit}^{-1}$}&&\multicolumn{2}{c}{$\text{cloglog}^{-1}$}\\
\cmidrule{5-6}\cmidrule{8-9}\cmidrule{11-12}
AUC&Method&$N_0=N_1$& & \multicolumn{1}{c}{Bias}&\multicolumn{1}{c}{RMSE}&&\multicolumn{1}{c}{Bias}&\multicolumn{1}{c}{RMSE}&&\multicolumn{1}{c}{Bias}&\multicolumn{1}{c}{RMSE}\\
\midrule
0.5           &\code{BoxCox}&25             && 0.001  & 0.083  && 0.000  & 0.082  && 0.000  & 0.082 \\
              &              &50             && 0.000  & 0.057  && 0.000  & 0.057  && 0.001  & 0.057 \\
              &              &100            && 0.000  & 0.041  && 0.000  & 0.040  && 0.000  & 0.040 \\
              &\code{Colr}  &25             && 0.001  & 0.084  && 0.000  & 0.083  && 0.001  & 0.083 \\
              &              &50             && 0.000  & 0.058  && -0.001 & 0.058  && 0.001  & 0.058 \\
              &              &100            && 0.000  & 0.041  && 0.000  & 0.041  && 0.000  & 0.041 \\
              &\code{Coxph} &25             && 0.001  & 0.077  && 0.000  & 0.076  && 0.000  & 0.077 \\
              &              &50             && 0.000  & 0.052  && 0.000  & 0.052  && 0.001  & 0.052 \\
              &              &100            && 0.000  & 0.036  && 0.000  & 0.036  && 0.000  & 0.036 \\
0.65          &\code{BoxCox}&25             && 0.003  & 0.078  && -0.002 & 0.078  && 0.004  & 0.078 \\
              &              &50             && 0.001  & 0.055  && -0.004 & 0.055  && 0.002  & 0.054 \\
              &              &100            && 0.000  & 0.038  && -0.006 & 0.040  && 0.002  & 0.038 \\
              &\code{Colr}  &25             && 0.001  & 0.078  && 0.002  & 0.078  && 0.002  & 0.079 \\
              &              &50             && 0.000  & 0.055  && 0.001  & 0.055  && 0.000  & 0.055 \\
              &              &100            && 0.000  & 0.038  && 0.000  & 0.039  && 0.001  & 0.038 \\
              &\code{Coxph} &25             && -0.021 & 0.077  && -0.028 & 0.080  && 0.006  & 0.071 \\
              &              &50             && -0.025 & 0.057  && -0.032 & 0.061  && 0.003  & 0.049 \\
              &              &100            && -0.027 & 0.044  && -0.034 & 0.050  && 0.002  & 0.034 \\
0.8           &\code{BoxCox}&25             && 0.004  & 0.061  && -0.002 & 0.066  && 0.003  & 0.063 \\
              &              &50             && 0.002  & 0.043  && -0.004 & 0.046  && 0.001  & 0.044 \\
              &              &100            && 0.001  & 0.030  && -0.006 & 0.033  && 0.000  & 0.031 \\
              &\code{Colr}  &25             && 0.000  & 0.061  && 0.002  & 0.063  && 0.002  & 0.062 \\
              &              &50             && -0.002 & 0.043  && 0.001  & 0.044  && 0.000  & 0.044 \\
              &              &100            && -0.002 & 0.030  && 0.000  & 0.031  && 0.000  & 0.031 \\
              &\code{Coxph} &25             && -0.038 & 0.075  && -0.049 & 0.086  && 0.007  & 0.054 \\
              &              &50             && -0.043 & 0.063  && -0.056 & 0.075  && 0.003  & 0.038 \\
              &              &100            && -0.047 & 0.057  && -0.060 & 0.070  && 0.001  & 0.027 \\
0.95          &\code{BoxCox}&25             && 0.001  & 0.027  && 0.003  & 0.030  && 0.003  & 0.031 \\
              &              &50             && 0.001  & 0.019  && 0.002  & 0.022  && 0.003  & 0.022 \\
              &              &100            && 0.001  & 0.014  && 0.002  & 0.016  && 0.003  & 0.016 \\
              &\code{Colr}  &25             && -0.006 & 0.028  && 0.000  & 0.028  && 0.001  & 0.028 \\
              &              &50             && -0.007 & 0.021  && 0.000  & 0.020  && 0.001  & 0.019 \\
              &              &100            && -0.007 & 0.015  && 0.000  & 0.014  && 0.001  & 0.014 \\
              &\code{Coxph} &25             && -0.038 & 0.054  && -0.042 & 0.065  && 0.003  & 0.023 \\
              &              &50             && -0.041 & 0.050  && -0.049 & 0.062  && 0.002  & 0.016 \\
              &              &100            && -0.044 & 0.049  && -0.054 & 0.061  && 0.002  & 0.011 \\
\bottomrule
\end{tabular}
 \caption{Bias and RMSE simulation results for estimation of the AUC. Data were generated using $F_0(y) = \Phi(y)$ for nondiseased results and $F_1(y) = \FZ( \FZinv(\Phi(y)) - \eshiftparm)$ for the diseased results. We varied $\FZ \in \{\probit^{-1}$, $\logit^{-1}$ and  $\cloglog^{-1} \}$, AUC and sample size. The proposed linear transformation models \code{BoxCox}, \code{Colr} and \code{Coxph} were used from the \pkg{tram} package to estimate $\eshiftparm$ and the transformation function $h=\FZinv \circ \Phi$.}
\label{tab:rmse}
\end{table}

\begin{table}\centering
\begin{tabular}{llllccc}
\toprule
&& && \multicolumn{3}{c}{$\FZ$}\\
\cmidrule{5-5}\cmidrule{6-6}\cmidrule{7-7}
AUC&Method&$N_0=N_1$ && \multicolumn{1}{c}{$\text{probit}^{-1}$}&\multicolumn{1}{c}{$\text{logit}^{-1}$}&\multicolumn{1}{c}{$\text{cloglog}^{-1}$}\\
\midrule
0.5           &\code{BoxCox}&25             && 0.888 & 0.890 & 0.890\\
              &              &50             && 0.895 & 0.894 & 0.894\\
              &              &100            && 0.896 & 0.900 & 0.894\\
              &\code{Colr}  &25             && 0.890 & 0.890 & 0.893\\
              &              &50             && 0.898 & 0.894 & 0.895\\
              &              &100            && 0.895 & 0.901 & 0.896\\
              &\code{Coxph} &25             && 0.874 & 0.875 & 0.873\\
              &              &50             && 0.887 & 0.887 & 0.886\\
              &              &100            && 0.895 & 0.897 & 0.886\\
0.65          &\code{BoxCox}&25             && 0.883 & 0.882 & 0.890\\
              &              &50             && 0.891 & 0.891 & 0.891\\
              &              &100            && 0.899 & 0.888 & 0.897\\
              &\code{Colr}  &25             && 0.887 & 0.891 & 0.894\\
              &              &50             && 0.894 & 0.897 & 0.892\\
              &              &100            && 0.903 & 0.895 & 0.895\\
              &\code{Coxph} &25             && 0.851 & 0.831 & 0.878\\
              &              &50             && 0.825 & 0.794 & 0.890\\
              &              &100            && 0.774 & 0.713 & 0.897\\
0.8           &\code{BoxCox}&25             && 0.892 & 0.868 & 0.883\\
              &              &50             && 0.893 & 0.876 & 0.885\\
              &              &100            && 0.898 & 0.874 & 0.893\\
              &\code{Colr}  &25             && 0.902 & 0.893 & 0.895\\
              &              &50             && 0.905 & 0.898 & 0.898\\
              &              &100            && 0.908 & 0.899 & 0.906\\
              &\code{Coxph} &25             && 0.762 & 0.685 & 0.892\\
              &              &50             && 0.650 & 0.538 & 0.892\\
              &              &100            && 0.462 & 0.304 & 0.896\\
0.95          &\code{BoxCox}&25             && 0.885 & 0.831 & 0.810\\
              &              &50             && 0.895 & 0.840 & 0.832\\
              &              &100            && 0.894 & 0.839 & 0.830\\
              &\code{Colr}  &25             && 0.907 & 0.888 & 0.867\\
              &              &50             && 0.901 & 0.897 & 0.887\\
              &              &100            && 0.877 & 0.896 & 0.889\\
              &\code{Coxph} &25             && 0.557 & 0.532 & 0.855\\
              &              &50             && 0.355 & 0.312 & 0.878\\
              &              &100            && 0.119 & 0.100 & 0.871\\
\bottomrule
\end{tabular}
 \caption{Observed coverage probabilities of nominal 90\% score confidence intervals for the AUC. Score intervals were calculated for $\eshiftparm$ from the proposed linear transformation models \code{BoxCox}, \code{Colr} and \code{Coxph}. The score intervals for the AUC were then calculated by transforming the intervals for $\eshiftparm$ using the closed-form expressions given in Table~\ref{tab:sum_simple}.}
\label{tab:cvg}
\end{table} 
\section{Application}
\label{sec:application}

The prevalence of obesity has increased consistently for most countries in the
recent decade and this trend is a serious global health concern
\citep{abarca2017worldwide}. Obesity contributes directly to increased risk of
cardiovascular disease (CVD) and its risk factors, including type 2 diabetes,
hypertension and dyslipidemia \citep{zalesin2008impact, grundy2004obesity}.
Metabolic syndrome (MetS) refers to the joint presence of several cardiovascular
risk factors and is characterized by insulin resistance
\citep{eckel2010metabolic}. The National Cholesterol Education Program Adult
Treatment Panel III (NCEP-ATP III) criteria is the
most widely used definition for MetS, but it requires laboratory analysis of a
blood sample. This has led to the search for non-invasive techniques which allow
reliable and early detection of MetS.

Waist-to-height ratio (WHtR) is a well-known anthropometric index used to
predict visceral obesity. Visceral obesity is an independent risk factor for
development of MetS by means of the increased production of free fatty acids
whose presence obstructs insulin activity \citep{bosello2000visceral}. This
suggests that higher values of WHtR, reflecting obesity and CVD risk factors,
are more indicative of incident MetS. Several studies have found that WHtR is
highly predictive of MetS \citep{shao2010waist, romero2016new,
suliga2019usefulness}. However, as waist circumference changes with age and
gender \citep{stevens2010associations}, it is also important to study whether or
not the performance of WHtR at diagnosing MetS is impacted by these variables.
Evaluation of WHtR as a predictor of MetS after adjusting for covariates is
necessary so that more tailored interventions can be initiated to improve
outcomes.

We illustrate the use of our methods to data from a cross sectional study
designed to validate the use of WHtR and systolic blood pressure (SBP) as
markers for early detection of MetS in a working population from the Balearic
Islands (Spain). Detailed descriptions of the study methodology and population
characteristics are reported in 
\AYcite{Romero-Salda{\~n}a et al.}{romero2018validation}. Briefly, data on
\numprint{60799} workers were collected during their work health periodic
assessments between 2012 and 2016. Presence of MetS was determined by the
NCEP-ATP III criteria and the sample consisted of \numprint{5487} workers with
MetS.

\subsection{Two-sample analysis}

\begin{table}[ht]
\centering
\begin{tabular}{cccc}
  \toprule 
 $\FZ$ & $\eshiftparm$ & AUC & $J$ \\
 \midrule 
 $\Phi$ & 1.492 (1.462, 1.521) & 0.854 (0.849, 0.859) & 0.544 (0.535, 0.553) \\ 
  $\expit$ & 2.785 (2.730, 2.841) & 0.871 (0.866, 0.875) & 0.602 (0.593, 0.611) \\ 
  $\text{cloglog}^{-1}$ & 1.186 (1.157, 1.215) & 0.766 (0.761, 0.771) & 0.412 (0.403, 0.421) \\ 
  $\text{loglog}^{-1}$ & 1.425 (1.397, 1.453) & 0.806 (0.802, 0.810) & 0.484 (0.475, 0.492) \\ 
   \bottomrule 
\end{tabular}
\caption{Estimates and 95\% score confidence intervals of the shift paramater, AUC and $J$ in the two-sample linear transformation model for WHtR as a marker of MetS.} 
\label{tab:comp}
\end{table}

\begin{figure}[t!]
\centering
\begin{knitrout}
\definecolor{shadecolor}{rgb}{0.969, 0.969, 0.969}\color{fgcolor}
\includegraphics[width=\linewidth]{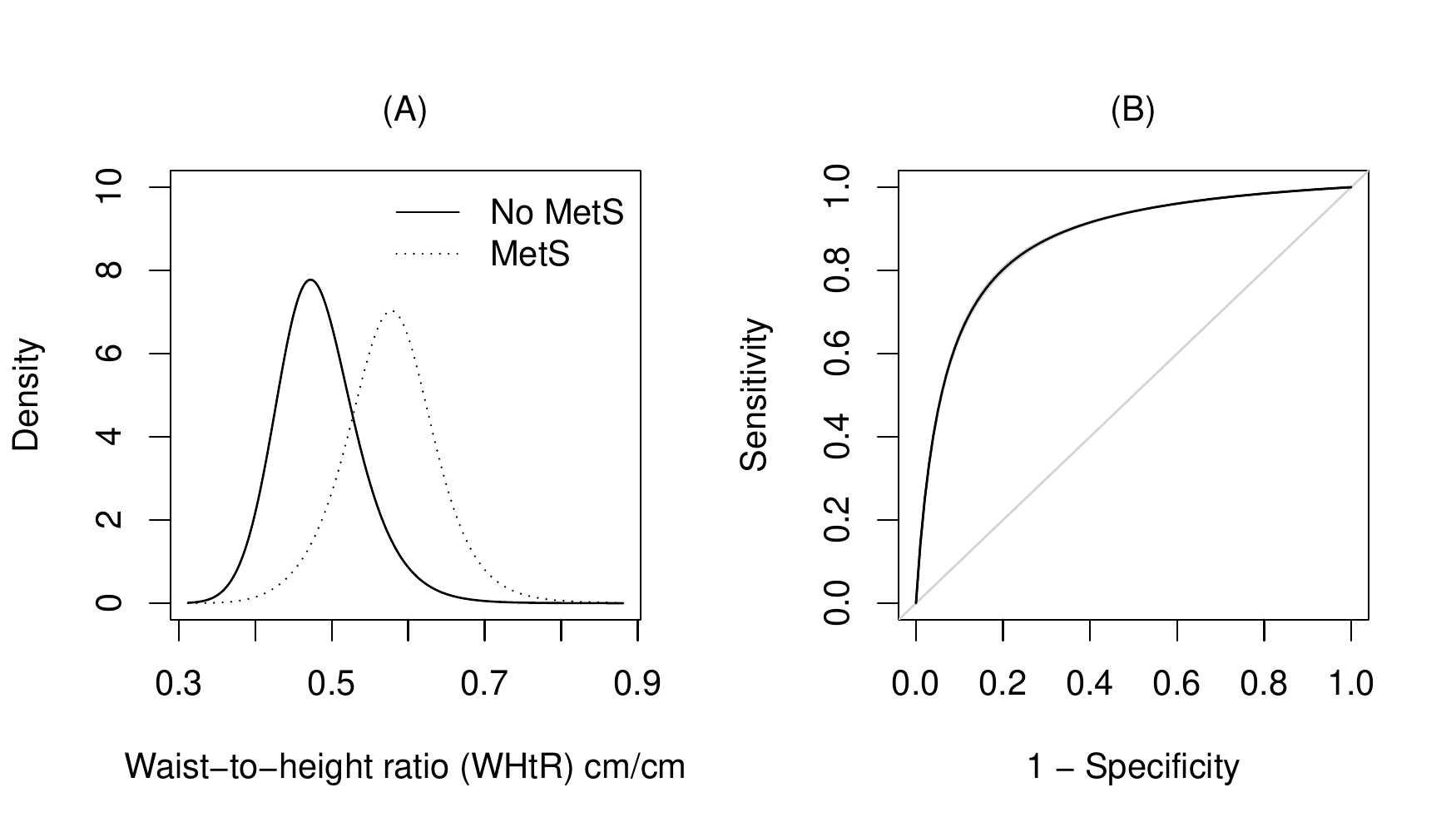} 
\end{knitrout}
\caption{Estimates from the linear transformation model with a single shift
parameter, $h(Y)=\eshiftparm d + Z$ where $Z$ is chosen to be a standard
logistic distribution. (A) Density functions of WHtR for the workers who were
diagnosed with MetS (dotted line) and those who were not (solid line). (B) ROC
curve for WHtR as a marker of MetS with 95\% uniform score confidence bands are
represented by gray shaded areas.}
\label{fig:ts}
\end{figure}

We first examined the unconditional performance of WHtR as a marker to diagnose
MetS, denoted $Y$ and $D$, respectively. We fit a linear transformation model
which leads to the ROC curve of the form in Equation~\ref{eq:roc_simple}, where
$\eshiftparm$ is the shift parameter, for various choices of the inverse link
function $\FZ$. Associated inference of the AUC and $J$ are calculated using the
closed-form expressions from Table~\ref{tab:sum_simple}. The resulting estimates
are presented in Table~\ref{tab:comp}. The AUCs are consistently bounded away
from 0.5 indicating a good capacity of WHtR to discriminate between workers with
and without MetS. This can also be seen from the estimated ROC curve plotted in
Figure~\ref{fig:ts} which lies well above the diagonal line as well as the
modeled densities which have a small degree of overlap. The confidence intervals
and uniform confidence bands are quite small due to the large sample size.

\subsection{Conditional ROC analysis}

\begin{figure}[t!]
\centering
\begin{knitrout}
\definecolor{shadecolor}{rgb}{0.969, 0.969, 0.969}\color{fgcolor}
\includegraphics[width=\linewidth]{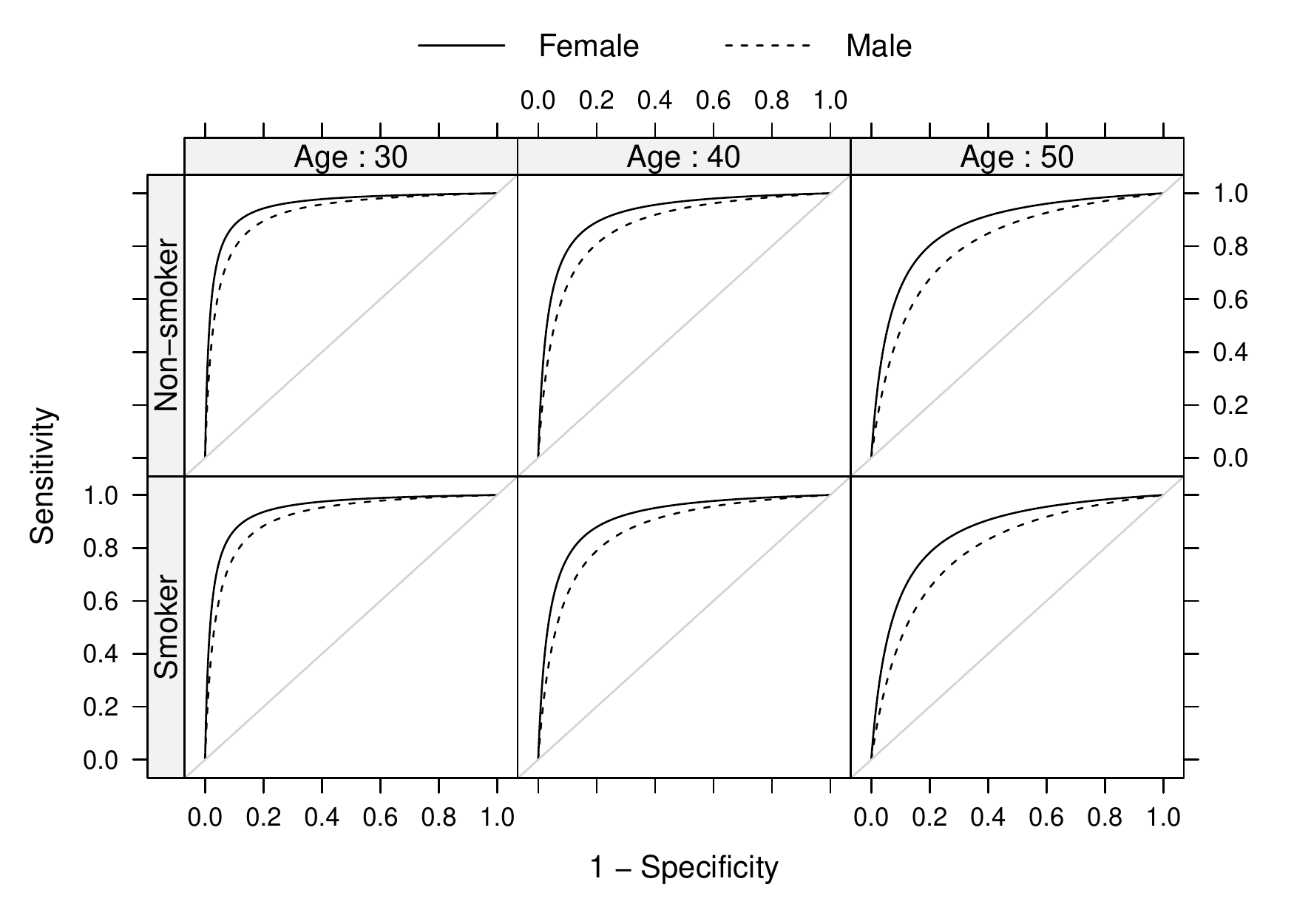} 
\end{knitrout}
\caption{Estimated covariate-specific ROC curves for WHtR as a marker of MetS
for female (solid line) and male workers (dashed line). Vertical panels
represent a specific age~(30, 40, 50) and horizontal panels smoking status.}
\label{fig:rage}
\end{figure}

\begin{figure}[t!]
\centering
\begin{knitrout}
\definecolor{shadecolor}{rgb}{0.969, 0.969, 0.969}\color{fgcolor}
\includegraphics[width=\linewidth]{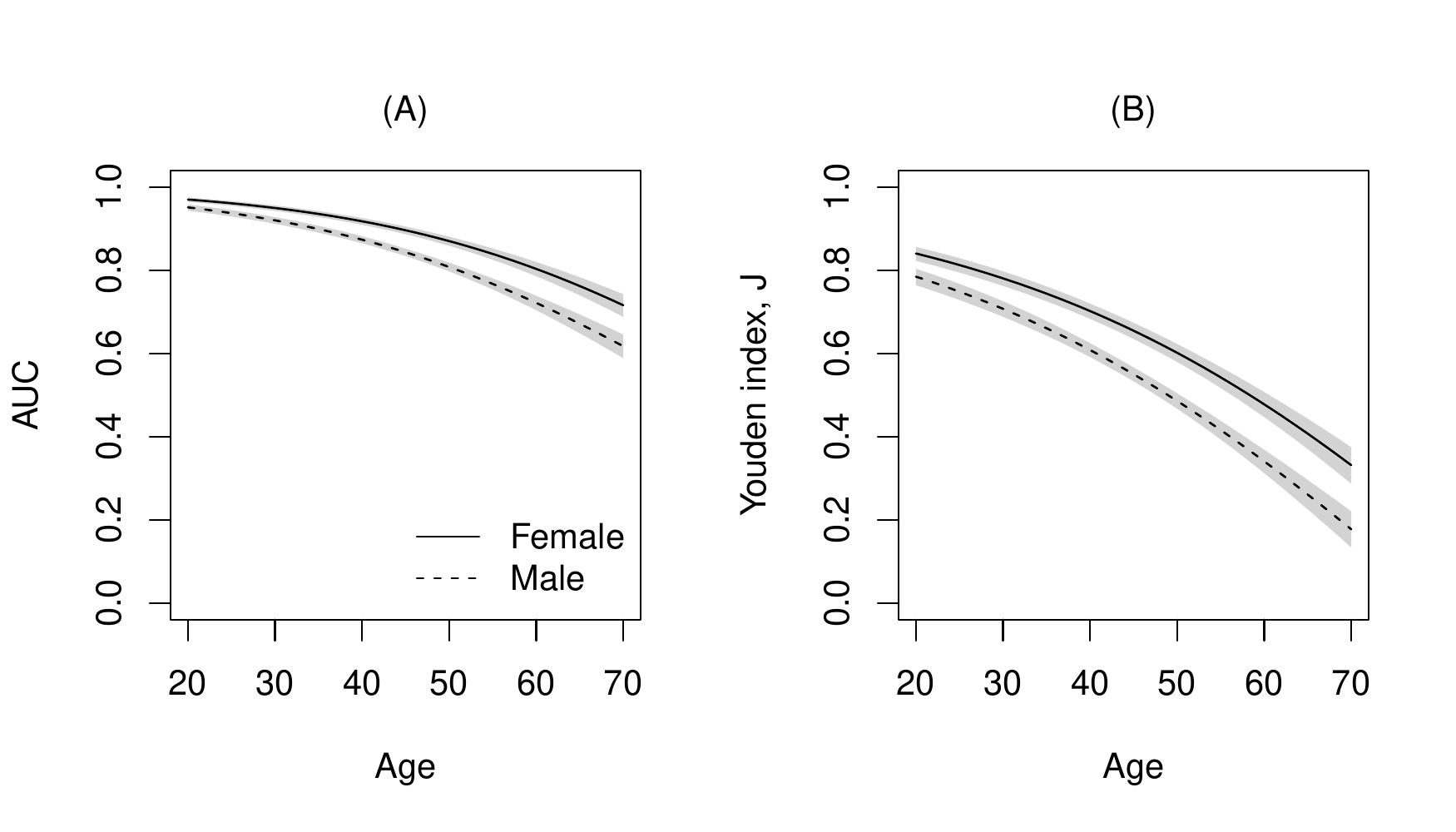} 
\end{knitrout}
\caption{Age-based AUC and Youden indices where WHtR is used as a marker to
detect MetS for non-smoking female (solid line) and male (dashed line) workers.
95\% Wald pointwise confidence bands are represented by gray shaded areas.}
\label{fig:auc}
\end{figure}

Next, we investigated if the discriminatory ability of WHtR in separating
workers with and without MetS varies with covariates. We considered a
transformation model that included the main effects of covariates plus
interaction terms with the disease indicator, which leads to the ROC curve given
by
\begin{align*}
\ROC(p \mid \rx) = 1 - \FZ(\FZinv(1-p) - (\eshiftparm + \gamma^\top \rx ))
\end{align*}
where the covariates $\rx$ included age, gender and tobacco consumption
and the inverse link function $\FZ=\logit^{-1}$. Figure~\ref{fig:rage} displays the
covariate-specific ROC curves fitted to these data. The performance of WHtR
appeared to be better for females compared to males and decreased with age. The
effect of smoking, although significant in the model, does not seem to
substantially alter the ROC curves given the other covariates are kept fixed. To
inspect the covariate effect further, we calculated the age- and gender-specific
AUCs and Youden indices from the model. Figure~\ref{fig:auc} clearly shows that
the discriminatory capabilities of WHtR in distinguishing workers with MetS is
consistently better for females and decreases with age.

\subsection{Comparing correlated biomarkers}

\begin{table}[ht]
\centering
\begin{tabular}{cccc}
  \toprule 
 Type & WHtR & SBP & Difference \\
 \midrule 
 $\eshiftparm_j$ & 1.874 (1.811, 1.945) & 1.214 (1.151, 1.273) & 0.660 (0.578, 0.750) \\ 
  AUC & 0.907 (0.900, 0.915) & 0.805 (0.792, 0.816) & 0.103 (0.090, 0.117) \\ 
  $J$ & 0.651 (0.635, 0.669) & 0.456 (0.435, 0.476) & 0.195 (0.171, 0.222) \\ 
   \bottomrule 
\end{tabular}
\caption{Marginal estimates and 95\% Wald confidence intervals of the shift paramater, AUC and Youden index from the two-sample bivariate normal transformation model for WHtR and SBP as markers of MetS for a 40-year non-smoking female worker. Wald confidence intervals are calculated by simulating from the asymptotic distribution of the MLE.} 
\label{tab:mult}
\end{table}

\begin{figure}[t!]
\begin{knitrout}
\definecolor{shadecolor}{rgb}{0.969, 0.969, 0.969}\color{fgcolor}

{\centering \includegraphics[width=4in]{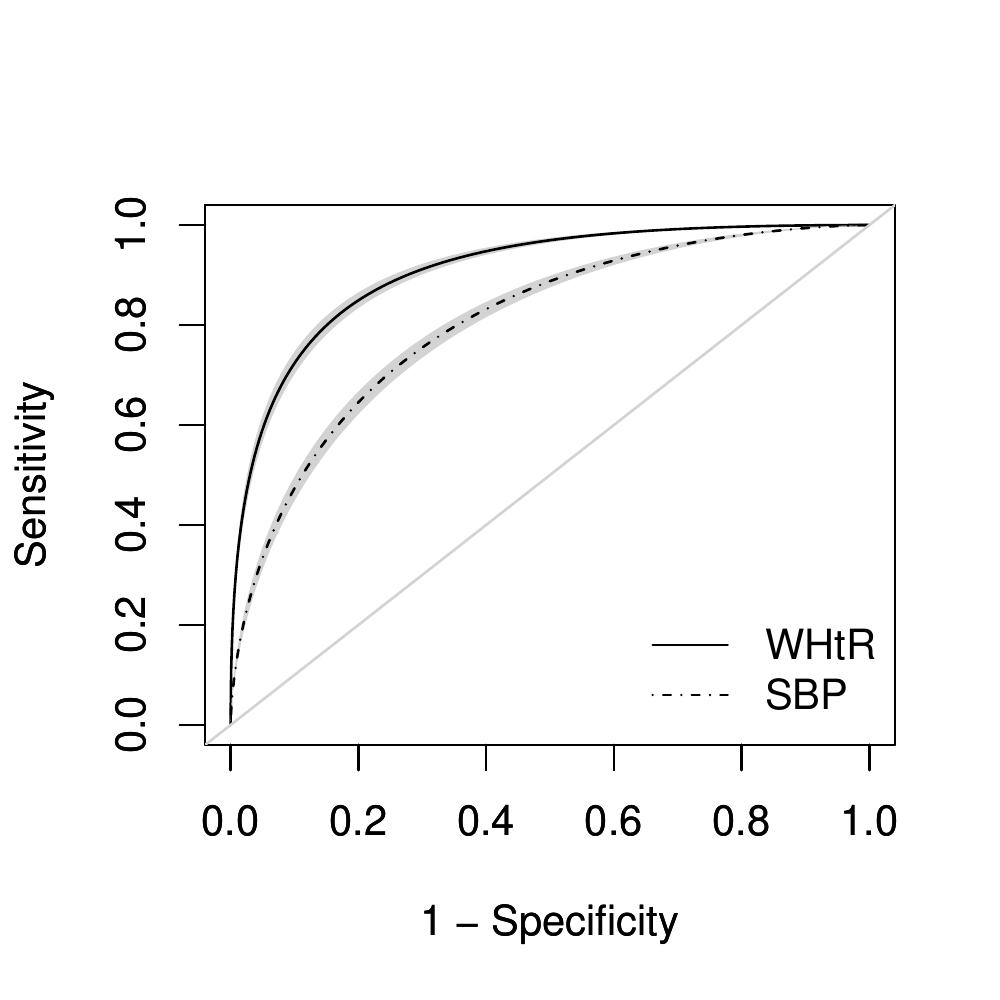} 

}

\end{knitrout}
\caption{Estimated marginal ROC curves for WHtR (solid line) and SBP (dot dashed line) as markers of MetS for a 40-year old non-smoking female worker. 95\% uniform confidence bands are represented by gray shaded areas.}
\label{fig:mauc}
\end{figure}

Finally, we compare the performance of WHtR to SBP, explicitly accounting for
the within-patient clustered nature of these measurements. That is, both markers
are measured on the same subject and hence their measurements are correlated. We
consider a transformation model for the joint distribution of $\mY=(\text{WHtR},
\text{SBP})$ that induces a marginal ROC curve for the $j$th marker, given by
\begin{align*}
\ROC_j(p \mid \rx) = 1 - \FZ(\FZinv(1-p) - (\eshiftparm_j + \gamma_j^\top \rx )) \quad \text{for } j \in \{\text{WHtR, SBP} \},
\end{align*}
where $\eshiftparm_j$ is the marginal shift parameter, $\gamma_j$ are the
marginal interaction coefficients and $\FZ=\probit^{-1}$. Figure~\ref{fig:mauc} shows
the corresponding estimated marginal ROC curves for the two markers for a
40-year old non-smoking female worker. The uniform confidence band is generated
by the simulation procedure detailed in Section~\ref{sec:simints}. Whilst both
markers perform well at predicting MetS, WHtR seems to be favorable with a ROC
curve uniformly above SBP. We tested the hypothesis for equality of ROC curves
$H_0: \ROC_1(p) = \ROC_2(p)$ for all $p \in [0,1]$, using the marginal
coefficients from the joint model. This hypothesis test is equivalent to testing
if the difference in the AUCs or Youden indices of the two markers is equal to
0. The estimates, differences and corresponding confidence intervals are
detailed in Table~\ref{tab:mult}. The difference of 0.103
between AUCs is significant ($p$~value $<0.001$) 
and indicates that the WHtR is better at discriminating incident 
MetS compared to SBP.

\section{Discussion}
This article presents a new modeling framework for ROC analysis that can be used to characterize the accuracy of medical diagnostic tests. Our model is based on estimating an unknown transformation function for the test results and yields a distribution-free yet model-based estimator for the ROC curve. Covariates that influence the diagnostic accuracy of tests can naturally be accommodated as regression parameters into the model and covariate-specific summary indices such as the AUC and Youden index are easily computed using closed-form expressions.
We extend the same conceptual transformation framework to a multivariate setting for comparing correlated diagnostic tests after appropriately accounting for their dependency structure. 

Our proposed approach has several features which distinguish it from contemporary methods of ROC analysis. Firstly, we employ maximum likelihood to jointly estimate all parameters defining the transformation function and regression coefficients. This implies the variation in the estimated transformation parameters is accounted for and appropriately propagated for the inference of the ROC curve. In turn, asymptotic efficiency is guaranteed for our estimators and we avoid reliance on resampling procedures for the construction of confidence intervals. Secondly, transformation models focus on estimating the conditional distribution function whose evaluation directly provides the likelihood contributions for interval, right- and left-censored data that commonly arises due to instrument detection limits. Additionally, with a simple reparameterization of the transformation function, our model extends to ROC curves for ordinal test results.
Thirdly, no strong assumptions are made regarding the transformation function which results in a highly flexible model that retains interpretability of the regression coefficients.
Fourthly, we develop a test for covariate-specific correlated ROC curves, a problem which has yet to be addressed in the literature.
Lastly, software implementations for all the methods described in this article are available in the \pkg{tram} \proglang{R} package (see Supplementary Material for example code), thus enabling a unified framework for ROC analysis.

One aspect that warrants further investigation is model selection, specifically with regards to the choice of $\FZ$. In general, this may depend on the aim of the study and distinct features of the data. 
Our simulation explored the extent to which the bias and coverage of the resulting estimates were affected by model choice. We found that a model with $\FZ=\logit^{-1}$ provided accurate results even when it was misspecified for the true data generating process. 
This model also behaves very similarly to the semiparametric cumulative probability model \citep{tian2020empirical}, also known as a proportional odds ordinal logistic model, both which estimate a log-odds ratio $\eshiftparm$.
The equivalence of the transformation model's odds ratio to the MWW test statistic has been well studied \citep{wang2017equivalence}.
The MWW statistic has a bounded influence function and is robust to contaminations of the specified model \citep{hampel1974influence}.
Due to their equivalence, we hypothesize that the transformation model with $\FZ=\logit^{-1}$ is also endowed with the same robustness properties as the MWW and can be chosen when no \emph{a priori} model is known. However, further research is needed to develop suitable model verification and selection methods.

Our proposed estimators assume that the difference between the diseased and nondiseased distributions is described by a shift-term, $\mu_d$ on the scale of $\FZ$. A relaxation of this assumption would allow for separate transformation functions in each of the two groups. Namely, consider a model where the nondiseased results follow a distribution with the CDF $F_0(y)=\FZ(h_0(y))$ and the diseased with the CDF $F_1(y)=\FZ(h_1(y))$. Defining a new transformation function $r=h_1 \circ h_0^{-1} \circ \FZinv : [0,1] \mapsto \RR$, the smooth ROC curve with no shift assumptions is given by $\ROC(p) = 1 - \FZ(r(1-p))$. This model has more flexibility but sacrifices the properness property desirable for ROC curves.
Furthermore, care has to be taken in defining the correct likelihood contributions for accurate inference of this model as uncertainty enters from both transformation functions.

In future work we plan to pursue various extensions of transformation models for ROC analysis to consider (1) penalty terms for high-dimensional covariates~\citep{kook2021regularized}, (2) mixed effects for clustered observations \citep{tamasi2022tramme}, and (3) covariate-dependent transformation functions through forest-based machine learning methods \citep{hothorn2021predictive}.

\section*{Funding}
	Financial support by Swiss National Science Foundation, grant number
	200021\_184603, is gratefully acknowledged.

\bibliography{refs}

\clearpage

\makeatletter
\renewcommand\thetable{S\@arabic\c@table}
\renewcommand \thefigure{S\@arabic\c@figure}
\renewcommand{\theHtable}{S\arabic{Table}}
\renewcommand{\theHfigure}{S\arabic{figure}}
\makeatother
\setcounter{table}{0}
\setcounter{figure}{0}

\begin{appendix}
\section{Simulation study}
\label{sec:sim}

\begin{table}[h!] \centering
\resizebox{\columnwidth}{!}{
	\begin{tabular}{llcccccc} \toprule 
	\multirow{2}{*}{Reference} 							& \multirow{2}{*}{\proglang{R} package}	& \multicolumn{2}{c}{ROC} 	& \multicolumn{2}{c}{AUC}  	& \multicolumn{2}{c}{Youden index} 	\\ \cmidrule{3-8}
										& 						& Estimate 	& CB 		& Estimate & CI 	& Estimate & CI 	\\ \midrule
	\cite{hothorn2020most} 				& \pkg{tram} 			& \cmark 	& \cmark 	& \cmark & \cmark 	& \cmark & \cmark	\\
	\cite{harrell2001regression, rms}	& \pkg{rms} 			&  	&  	& \cmark & \cmark 	&  & 	\\
	\cite{thas2012probabilistic}		& \pkg{pim} 			&  	&  	& \cmark & \cmark 	&  & 	\\
	\cite{survival}				 		& \pkg{survival}		&  	&  	& \cmark & \cmark 	&  & 	\\
	\cite{robin2011proc}				& \pkg{pROC}			& \cmark 	& \cmark 	& \cmark & \cmark 	& \cmark & 	\\
	\cite{asht}							& \pkg{asht}			&  	&  	& \cmark & \cmark 	&  & 	\\
	\cite{konietschke2015nparcomp}		& \pkg{nparcomp} 		&  	&  	& \cmark & \cmark 	&  & 	\\
	\cite{rocit} 						& \pkg{ROCit} 			& \cmark 	& \cmark 	& \cmark & \cmark 	& \cmark & 	\\
	\cite{auroc} 						& \pkg{auRoc} 			&  	&  	& \cmark & \cmark 	&  & 	\\
	\cite{perez2017thresholdroc}  		& \pkg{ThresholdROC} 	&  	&  	&  &  	& \cmark & \cmark	\\
	\cite{ridout2009ovl} 				& \pkg{overlap} 		&  	&  	&  &  	& \cmark & \cmark	\\
	\cite{franco2021inference}  		& -						&  	&  	&  &  	& \cmark & \cmark	\\
	\cite{perez2018nsroc} 				& \pkg{nsROC} 			& \cmark 	& \cmark 	& \cmark & \cmark 	& \cmark & 	\\
	\bottomrule
	\end{tabular}
}
\caption{Overview of the different methods used in the simulation study. References to the original publication along with \proglang{R} software details are given. The (\cmark) indicates if a method computes the specific metric. The metrics included estimates for the ROC curve, AUC and Youden index as well as corresponding confidence bands or confidence intervals.}
\label{tab:comp_mthds}
\end{table}

\begin{figure}[h!]
	\centering
	\includegraphics[width=1\textwidth]{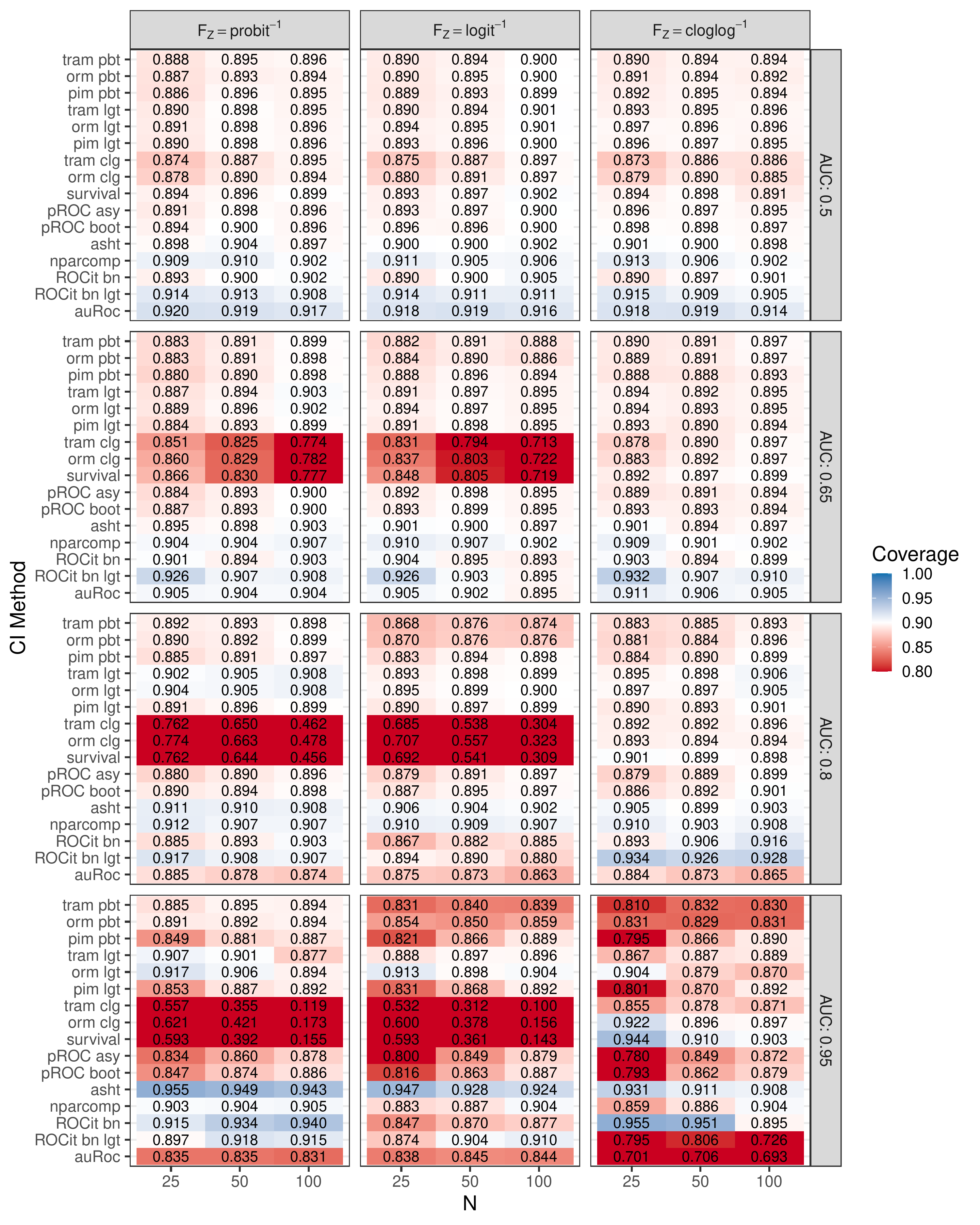}
	\caption{Empirical coverage probabilities of 90\% confidence intervals for the AUC from the different methods used in the simulation study. The methods are denoted by their \proglang{R}~package name from Table~\ref{tab:comp_mthds} followed by a short abbreviation to uniquely identify methods with multiple procedures. These included ``pbt'' ($\probit^{-1}$), ``lgt'' ($\logit^{-1}$), ``clg'' ($\cloglog^{-1}$), ``asy'' (asymptotic), ``boot'' (bootstrap) and ``bn'' (binormal).}
	\label{fig:cvg_auc}
\end{figure}

\begin{figure}[h!]
	\centering
	\includegraphics[width=1\textwidth]{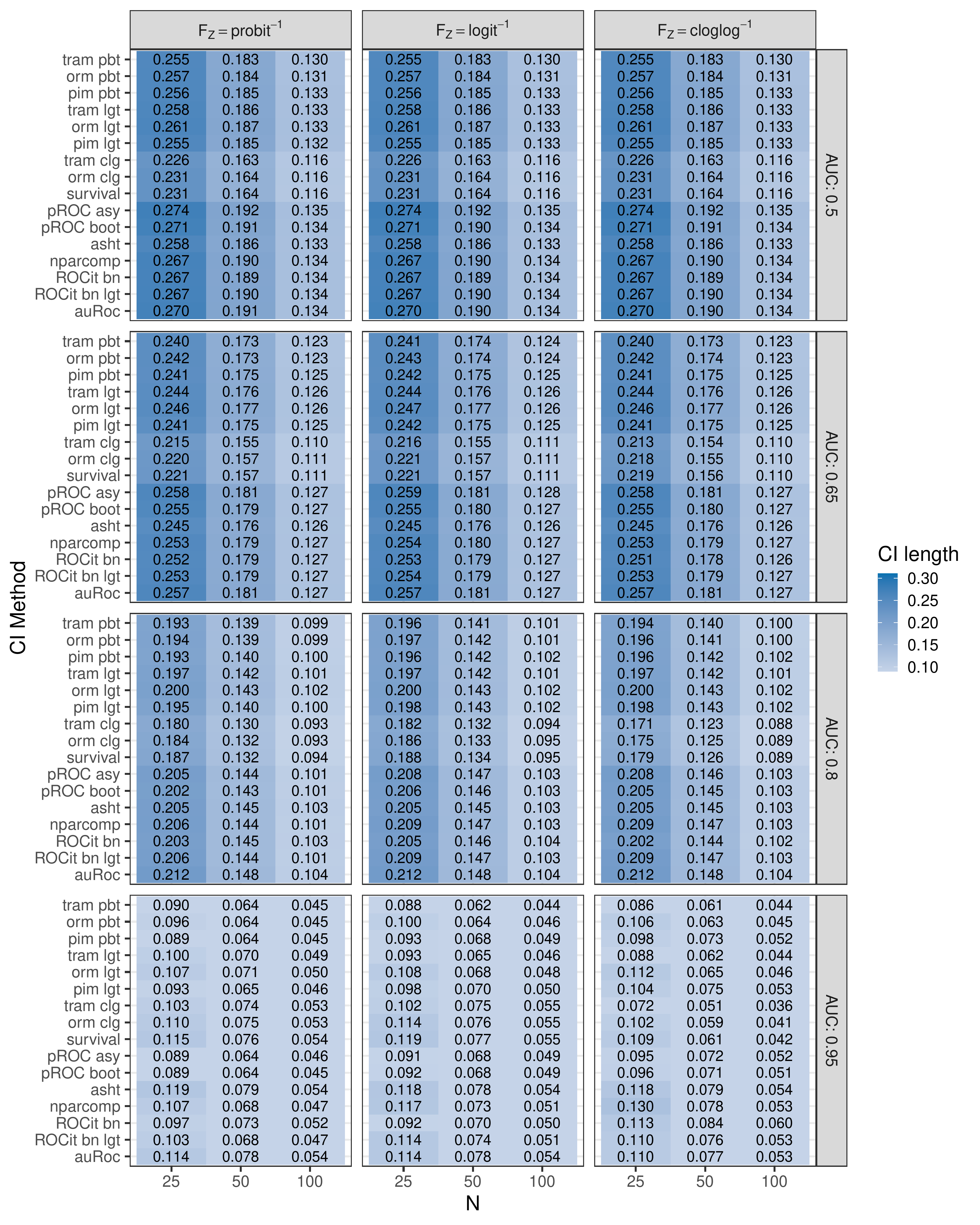}
	\caption{Average width of 90\% confidence intervals for the AUC from the different methods used in the simulation study. The methods are denoted by their \proglang{R}~package name from Table~\ref{tab:comp_mthds} followed by a short abbreviation to uniquely identify methods with multiple procedures. These included ``pbt'' ($\probit^{-1}$), ``lgt'' ($\logit^{-1}$), ``clg'' ($\cloglog^{-1}$), ``asy'' (asymptotic), ``boot'' (bootstrap) and ``bn'' (binormal).}
	\label{fig:pwr_auc}
\end{figure}

\begin{figure}[h!]
	\centering
	\includegraphics[width=1\textwidth]{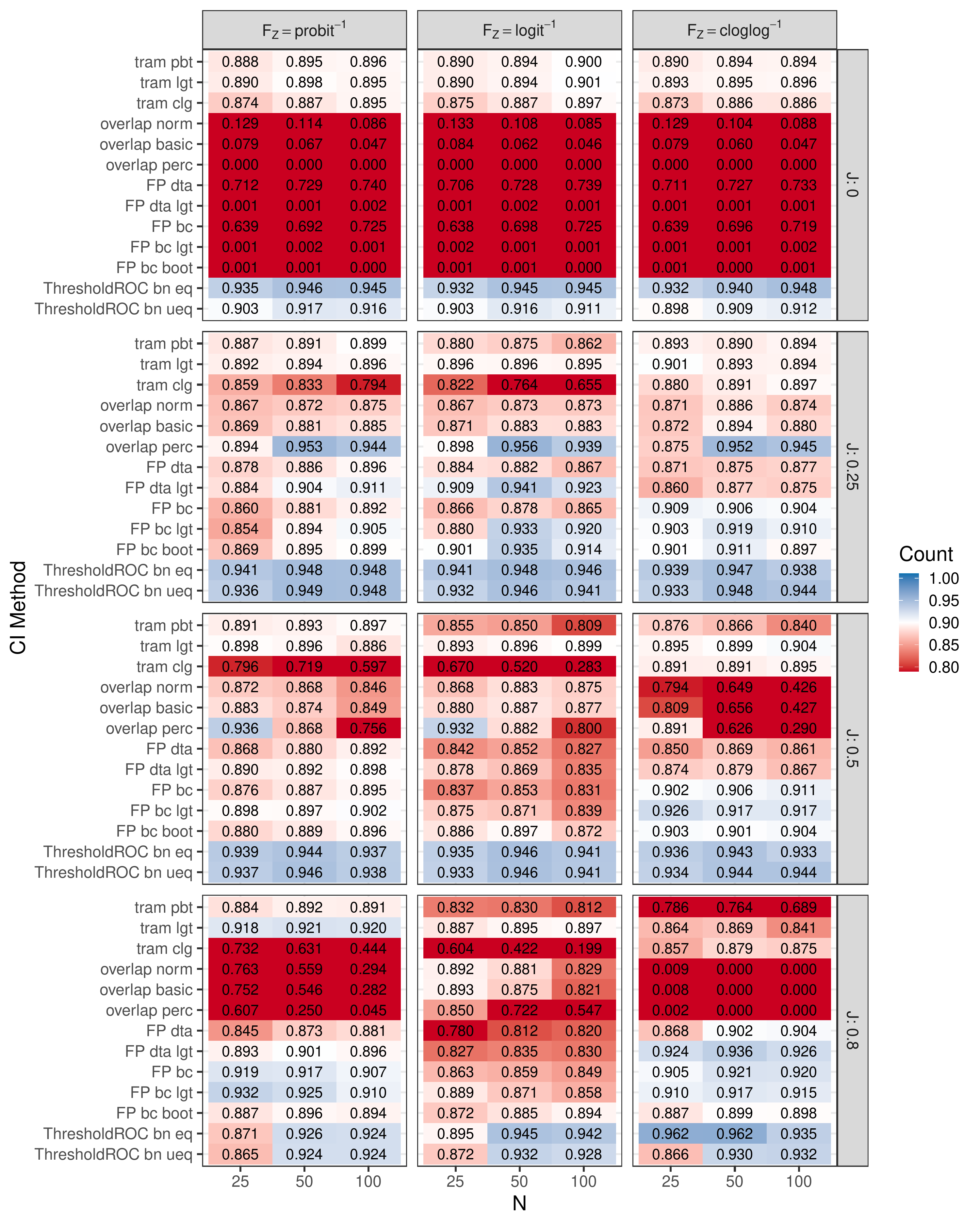}
	\caption{Empirical coverage probabilities of 90\% confidence intervals for the Youden index from the different methods used in the simulation study. The methods are denoted by their \proglang{R}~package name from Table~\ref{tab:comp_mthds} followed by a short abbreviation to uniquely identify methods with multiple procedures. These included ``pbt'' ($\probit^{-1}$), ``lgt'' ($\logit^{-1}$), ``clg'' ($\cloglog^{-1}$), ``norm'' (bootstrap normal approximation interval), ``basic'' (basic bootstrap interval), ``perc'' (bootstrap percentile interval), ``dta'' (delta method), ``bc'' (BoxCox transformation), ``boot'' (bootstrap variance), ``bn eq'' (binormal equal variance) and ``bn uneq'' (binormal unequal variance).}
	\label{fig:cvg_yi}
\end{figure}

\begin{figure}[h!]
	\centering
	\includegraphics[width=1\textwidth]{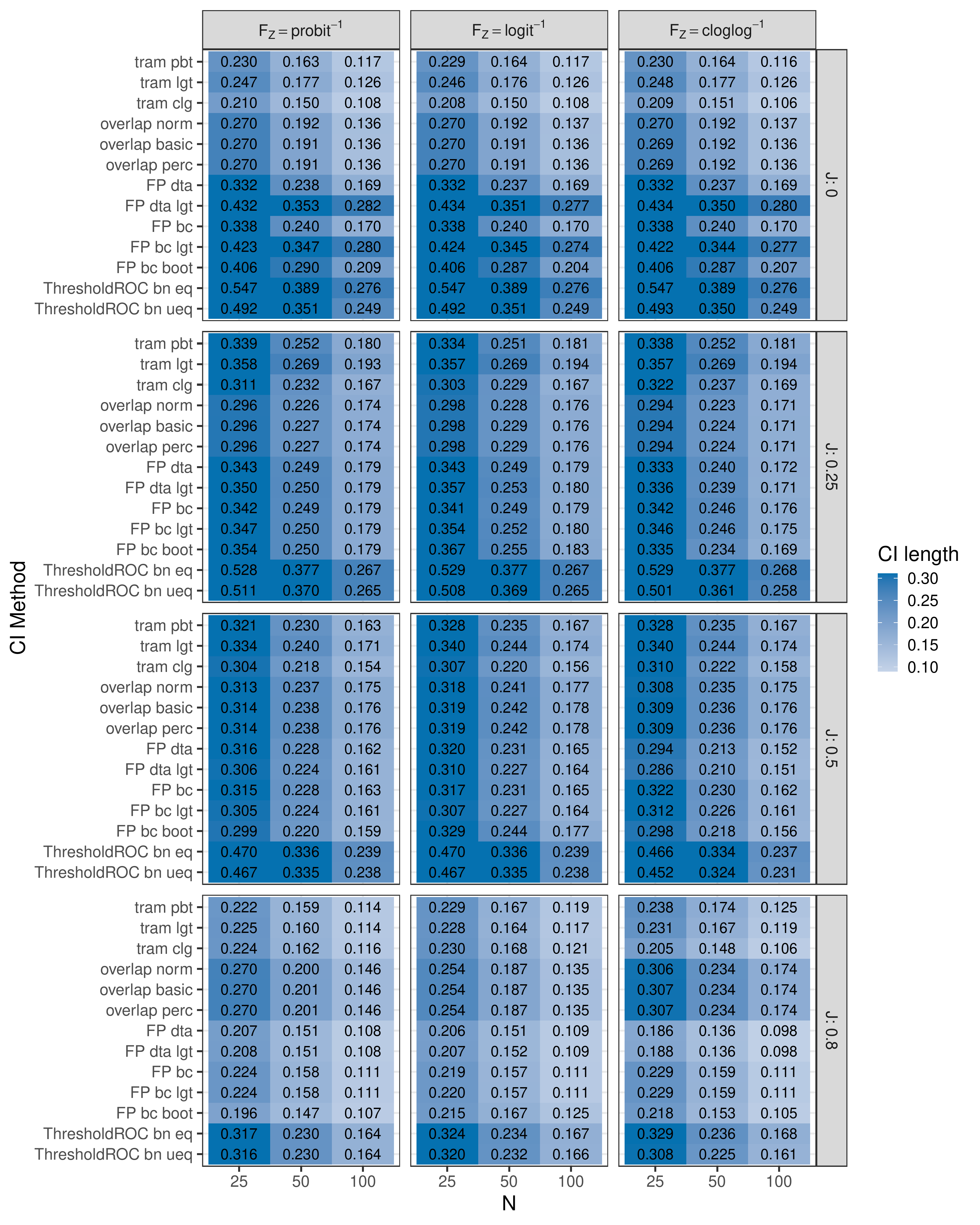}
	\caption{Average width of 90\% confidence intervals for the AUC from the different methods used in the simulation study. The methods are denoted by their \proglang{R}~package name from Table~\ref{tab:comp_mthds} followed by a short abbreviation to uniquely identify methods with multiple procedures. These included ``pbt'' ($\probit^{-1}$), ``lgt'' ($\logit^{-1}$), ``clg'' ($\cloglog^{-1}$), ``norm'' (bootstrap normal approximation interval), ``basic'' (basic bootstrap interval), ``perc'' (bootstrap percentile interval), ``dta'' (delta method), ``bc'' (BoxCox transformation), ``boot'' (bootstrap variance), ``bn eq'' (binormal equal variance) and ``bn uneq'' (binormal unequal variance).}
	\label{fig:pwr_yi}
\end{figure}

\begin{figure}[h!]
	\centering
	\includegraphics[width=1\textwidth]{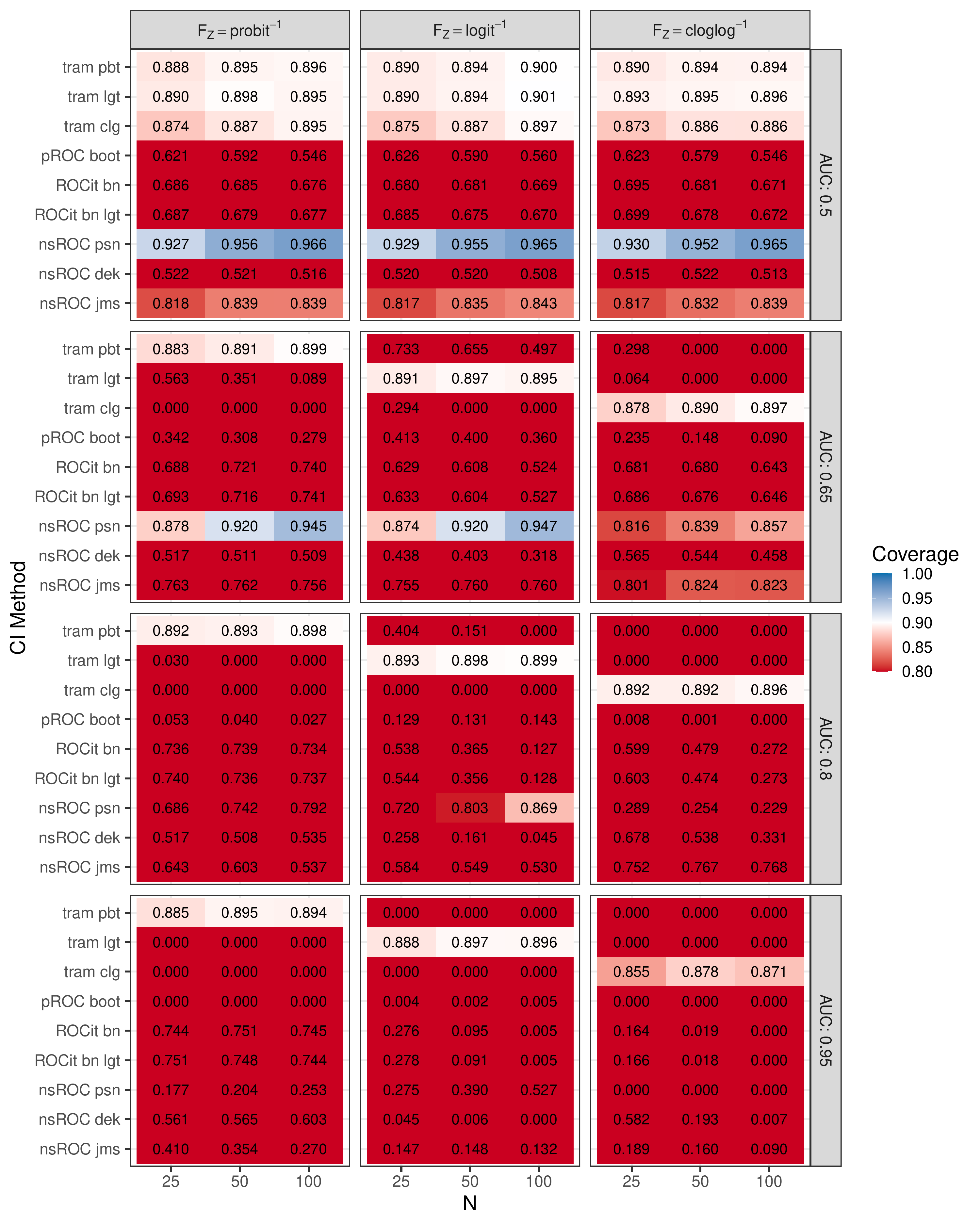}
	\caption{Empirical coverage probabilities of 90\% confidence bands for the ROC curve from the different methods used in the simulation study. The methods are denoted by their \proglang{R}~package name from Table~\ref{tab:comp_mthds} followed by a short abbreviation to uniquely identify methods with multiple procedures. These included ``pbt'' ($\probit^{-1}$), ``lgt'' ($\logit^{-1}$), ``clg'' ($\cloglog^{-1}$), ``boot'' (empirical bootstrap), ``bn'' (binormal), ``psn''~\citep{martinez2018efficient}, ``dek''~\citep{demidenko2012confidence} and ``jms''~\citep{jensen2000regional}.}
	\label{fig:cvg_roc}
\end{figure}

\begin{figure}[h!]
	\centering
	\includegraphics[width=1\textwidth]{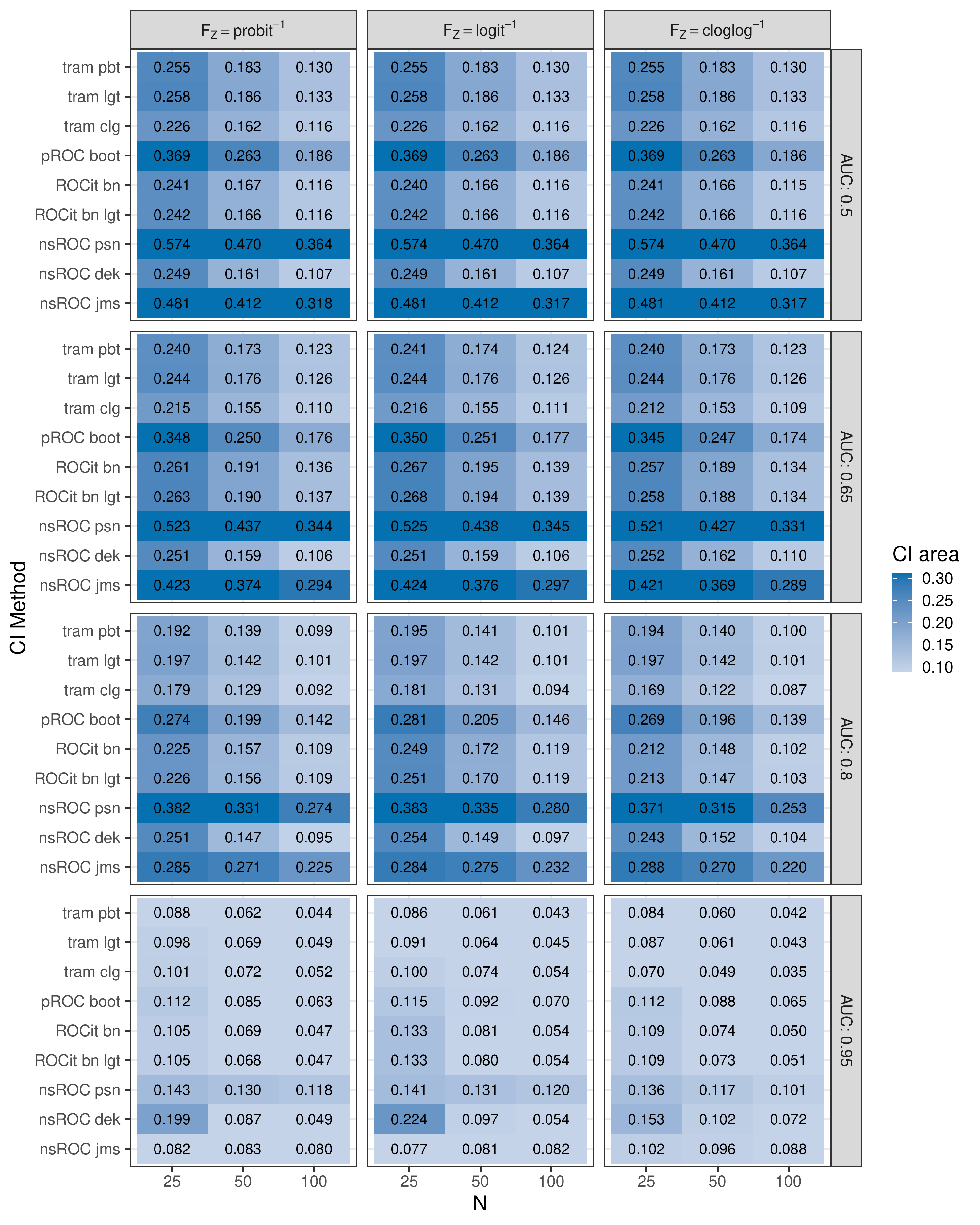}
	\caption{Average areas of 90\% confidence bands for the ROC curve from the different methods used in the simulation study. The methods are denoted by their \proglang{R}~package name from Table~\ref{tab:comp_mthds} followed by a short abbreviation to uniquely identify methods with multiple procedures. These included ``pbt'' ($\probit^{-1}$), ``lgt'' ($\logit^{-1}$), ``clg'' ($\cloglog^{-1}$), ``boot'' (empirical bootstrap), ``bn'' (binormal), ``psn''~\citep{martinez2018efficient}, ``dek''~\citep{demidenko2012confidence} and ``jms''~\citep{jensen2000regional}.}
	\label{fig:pwr_roc}
\end{figure} \end{appendix}

\clearpage

\section{Computational details}
\label{sec:software}

This section provides \proglang{R} code for fitting transformation models and conducting ROC analysis for the metabolic syndrome (MetS) application detailed in the paper. We first consider the unconditional performance of waist-to-height ratio (WHtR) as a diagnostic test of MetS. Next, we examined if the discriminatory ability of WHtR varies with covariates. Finally, we compare the performance of an alternative diagnostic test for MetS, systolic blood pressure (SBP). The structure of the code follows the tables and figures presented in the main text. 

\subsection{Two-sample analysis}
\begin{knitrout}
\definecolor{shadecolor}{rgb}{0.969, 0.969, 0.969}\color{fgcolor}\begin{kframe}
\begin{alltt}
\hlkwd{library}\hlstd{(}\hlstr{"tram"}\hlstd{)}
\hlkwd{library}\hlstd{(}\hlstr{"rdryad"}\hlstd{)}

\hlcom{### MetS data; doi: 10.5061/dryad.cb51t54}
\hlstd{dd} \hlkwb{<-} \hlkwd{dryad_files_download}\hlstd{(}\hlnum{79303}\hlstd{)}
\hlstd{dat} \hlkwb{<-} \hlkwd{read.csv}\hlstd{(dd[[}\hlnum{1}\hlstd{]])}
\hlstd{dat}\hlopt{$}\hlstd{D} \hlkwb{<-} \hlstd{dat}\hlopt{$}\hlstd{MetS}
\hlstd{dat}\hlopt{$}\hlstd{Y} \hlkwb{<-} \hlstd{dat}\hlopt{$}\hlstd{WHtR}
\hlstd{supp} \hlkwb{<-} \hlkwd{with}\hlstd{(dat,} \hlkwd{c}\hlstd{(}\hlkwd{min}\hlstd{(Y),} \hlkwd{max}\hlstd{(Y)))}
\hlstd{nd} \hlkwb{<-} \hlkwd{data.frame}\hlstd{(}\hlkwc{D} \hlstd{=} \hlkwd{c}\hlstd{(}\hlnum{0}\hlstd{,}\hlnum{1}\hlstd{))}
\hlstd{ndD1} \hlkwb{<-} \hlstd{nd[}\hlnum{2}\hlstd{,,} \hlkwc{drop} \hlstd{=} \hlnum{FALSE}\hlstd{]}

\hlcom{### Two-sample ROC analysis}
\hlcom{### Model with F_Z = logit^\{-1\}}
\hlstd{(m_lgt} \hlkwb{<-} \hlkwd{Colr}\hlstd{(Y} \hlopt{~} \hlstd{D,} \hlkwc{data} \hlstd{= dat,} \hlkwc{support} \hlstd{= supp))}

\hlcom{### ROC curve with 95 percent Wald CI}
\hlstd{r} \hlkwb{<-} \hlkwd{ROC}\hlstd{(m_lgt,} \hlkwc{newdata} \hlstd{= ndD1,} \hlkwc{conf.level} \hlstd{=} \hlnum{0.95}\hlstd{)}
\hlkwd{plot}\hlstd{(r)}
\hlcom{### Score test}
\hlstd{(sc} \hlkwb{<-} \hlkwd{score_test}\hlstd{(m_lgt,} \hlkwc{level} \hlstd{=} \hlnum{0.95}\hlstd{))}

\hlcom{### Summary indices and Wald CIs}
\hlkwd{PI}\hlstd{(m_lgt,} \hlkwc{newdata} \hlstd{= ndD1,} \hlkwc{conf.level} \hlstd{=} \hlnum{0.95}\hlstd{)}
\hlkwd{TV}\hlstd{(m_lgt,} \hlkwc{newdata} \hlstd{= ndD1,} \hlkwc{conf.level} \hlstd{=} \hlnum{0.95}\hlstd{)}
\hlkwd{OVL}\hlstd{(m_lgt,} \hlkwc{newdata} \hlstd{= ndD1,} \hlkwc{conf.level} \hlstd{=} \hlnum{0.95}\hlstd{)}

\hlcom{### Score CIs for summary indices}
\hlkwd{PI}\hlstd{(sc}\hlopt{$}\hlstd{conf.int,} \hlkwc{link} \hlstd{=} \hlstr{"logistic"}\hlstd{)}
\hlkwd{TV}\hlstd{(}\hlkwd{rev}\hlstd{(sc}\hlopt{$}\hlstd{conf.int),} \hlkwc{link} \hlstd{=} \hlstr{"logistic"}\hlstd{)}
\hlkwd{OVL}\hlstd{(sc}\hlopt{$}\hlstd{conf.int,} \hlkwc{link} \hlstd{=} \hlstr{"logistic"}\hlstd{)}

\hlcom{### Modeled densities}
\hlstd{grd} \hlkwb{<-} \hlkwd{seq}\hlstd{(supp[}\hlnum{1}\hlstd{], supp[}\hlnum{2}\hlstd{],} \hlkwc{by}\hlstd{=}\hlnum{0.001}\hlstd{)}
\hlstd{dens} \hlkwb{<-} \hlkwd{predict}\hlstd{(m_lgt, nd,} \hlkwc{type} \hlstd{=} \hlstr{"density"}\hlstd{,} \hlkwc{q}\hlstd{=grd)}
\hlkwd{plot}\hlstd{(grd, dens[,}\hlnum{1}\hlstd{],} \hlkwc{type} \hlstd{=} \hlstr{"l"}\hlstd{,} \hlkwc{ylim} \hlstd{=} \hlkwd{c}\hlstd{(}\hlnum{0}\hlstd{,} \hlnum{10}\hlstd{),}
     \hlkwc{xlab} \hlstd{=} \hlstr{"Waist-to-height ratio (WHtR) cm/cm"}\hlstd{,}
     \hlkwc{ylab} \hlstd{=} \hlstr{"Density"}\hlstd{)}
\hlkwd{lines}\hlstd{(grd, dens[,}\hlnum{2}\hlstd{],} \hlkwc{lty} \hlstd{=} \hlnum{3}\hlstd{)}
\hlkwd{legend}\hlstd{(}\hlstr{"topright"}\hlstd{,} \hlkwc{legend}\hlstd{=}\hlkwd{c}\hlstd{(}\hlstr{"No MetS"}\hlstd{,} \hlstr{"MetS"}\hlstd{),} \hlkwc{lty} \hlstd{=} \hlkwd{c}\hlstd{(}\hlnum{1}\hlstd{,} \hlnum{3}\hlstd{))}
\end{alltt}
\end{kframe}
\end{knitrout}

\subsection{Conditional ROC analysis}
\begin{knitrout}
\definecolor{shadecolor}{rgb}{0.969, 0.969, 0.969}\color{fgcolor}\begin{kframe}
\begin{alltt}
\hlstd{(m_cov} \hlkwb{<-} \hlkwd{Colr}\hlstd{(Y} \hlopt{~} \hlstd{D} \hlopt{*} \hlstd{(Age} \hlopt{+} \hlstd{Gender} \hlopt{+} \hlstd{Smoke),} \hlkwc{data} \hlstd{= dat,} \hlkwc{support} \hlstd{= supp))}

\hlstd{grd_age} \hlkwb{<-} \hlnum{20}\hlopt{:}\hlnum{70}
\hlstd{nd1} \hlkwb{<-} \hlkwd{expand.grid}\hlstd{(}\hlkwc{D} \hlstd{=} \hlnum{1}\hlstd{,} \hlkwc{Age} \hlstd{= grd_age,} \hlkwc{Gender} \hlstd{=} \hlkwd{c}\hlstd{(}\hlnum{0}\hlstd{,} \hlnum{1}\hlstd{),} \hlkwc{Smoke} \hlstd{=} \hlnum{0}\hlstd{)}
\hlstd{nd0} \hlkwb{<-} \hlkwd{expand.grid}\hlstd{(}\hlkwc{D} \hlstd{=} \hlnum{0}\hlstd{,} \hlkwc{Age} \hlstd{= grd_age,} \hlkwc{Gender} \hlstd{=} \hlkwd{c}\hlstd{(}\hlnum{0}\hlstd{,} \hlnum{1}\hlstd{),} \hlkwc{Smoke} \hlstd{=} \hlnum{0}\hlstd{)}

\hlcom{### Covariate-specific ROC curve for 40 year old non-smokers}
\hlcom{### Solid line: Female, Dashed line: Male}
\hlstd{p} \hlkwb{<-} \hlkwd{seq}\hlstd{(}\hlnum{0}\hlstd{,} \hlnum{1}\hlstd{,} \hlkwc{by} \hlstd{=} \hlnum{0.001}\hlstd{)}
\hlstd{rcov} \hlkwb{<-} \hlkwd{ROC}\hlstd{(m_cov,}
            \hlkwc{newdata} \hlstd{=} \hlkwd{subset}\hlstd{(nd1, Age} \hlopt{==} \hlnum{40}\hlstd{),}
            \hlkwc{reference} \hlstd{=} \hlkwd{subset}\hlstd{(nd0, Age} \hlopt{==} \hlnum{40}\hlstd{),}
            \hlkwc{one2one} \hlstd{= T,} \hlkwc{prob} \hlstd{= p)}
\hlkwd{plot}\hlstd{(rcov)}
\hlcom{### Age-based AUC for non-smokers}
\hlcom{### Solid line: Female, Dashed line: Male}
\hlstd{auc} \hlkwb{<-} \hlkwd{PI}\hlstd{(m_cov,}
          \hlkwc{newdata} \hlstd{= nd1,}
          \hlkwc{reference} \hlstd{= nd0,}
          \hlkwc{one2one} \hlstd{=} \hlnum{TRUE}\hlstd{,} \hlkwc{conf.level} \hlstd{=} \hlnum{0.95}\hlstd{)}
\hlstd{id_f} \hlkwb{<-} \hlkwd{which}\hlstd{(nd1}\hlopt{$}\hlstd{Gender}\hlopt{==}\hlnum{0}\hlstd{)}
\hlstd{id_m} \hlkwb{<-} \hlkwd{which}\hlstd{(nd1}\hlopt{$}\hlstd{Gender}\hlopt{==}\hlnum{1}\hlstd{)}

\hlcom{## Females}
\hlkwd{plot}\hlstd{(grd_age, auc[id_f,} \hlstr{"Estimate"}\hlstd{],}
     \hlkwc{xlab} \hlstd{=} \hlstr{"Age (in years)"}\hlstd{,} \hlkwc{ylab} \hlstd{=} \hlstr{"AUC"}\hlstd{,}
     \hlkwc{ylim} \hlstd{=} \hlkwd{c}\hlstd{(}\hlnum{0}\hlstd{,} \hlnum{1}\hlstd{),} \hlkwc{type} \hlstd{=} \hlstr{"l"}\hlstd{)}
\hlkwd{lines}\hlstd{(grd_age, auc[id_f,} \hlstr{"lwr"}\hlstd{],} \hlkwc{lty} \hlstd{=} \hlnum{3}\hlstd{,} \hlkwc{col} \hlstd{=} \hlstr{"gray50"}\hlstd{)}
\hlkwd{lines}\hlstd{(grd_age, auc[id_f,} \hlstr{"upr"}\hlstd{],} \hlkwc{lty} \hlstd{=} \hlnum{3}\hlstd{,} \hlkwc{col} \hlstd{=} \hlstr{"gray50"}\hlstd{)}

\hlcom{## Males}
\hlkwd{lines}\hlstd{(grd_age, auc[id_m,} \hlstr{"Estimate"}\hlstd{],} \hlkwc{lty} \hlstd{=} \hlnum{2}\hlstd{)}
\hlkwd{lines}\hlstd{(grd_age, auc[id_m,} \hlstr{"lwr"}\hlstd{],} \hlkwc{lty} \hlstd{=} \hlnum{3}\hlstd{,} \hlkwc{col} \hlstd{=} \hlstr{"gray50"}\hlstd{)}
\hlkwd{lines}\hlstd{(grd_age, auc[id_m,} \hlstr{"upr"}\hlstd{],} \hlkwc{lty} \hlstd{=} \hlnum{3}\hlstd{,} \hlkwc{col} \hlstd{=} \hlstr{"gray50"}\hlstd{)}
\hlkwd{legend}\hlstd{(}\hlstr{"bottomright"}\hlstd{,} \hlkwc{legend} \hlstd{=} \hlkwd{c}\hlstd{(}\hlstr{"Female"}\hlstd{,} \hlstr{"Male"}\hlstd{),} \hlkwc{lty} \hlstd{=} \hlnum{1}\hlopt{:}\hlnum{2}\hlstd{,} \hlkwc{bty} \hlstd{=} \hlstr{"n"}\hlstd{)}
\end{alltt}
\end{kframe}
\end{knitrout}

\subsection{Comparing correlated biomarkers}
\begin{knitrout}
\definecolor{shadecolor}{rgb}{0.969, 0.969, 0.969}\color{fgcolor}\begin{kframe}
\begin{alltt}
\hlkwd{library}\hlstd{(}\hlstr{"mvtnorm"}\hlstd{)}
\hlstd{dat}\hlopt{$}\hlstd{Y1} \hlkwb{<-} \hlstd{dat}\hlopt{$}\hlstd{WHtR}
\hlstd{dat}\hlopt{$}\hlstd{Y2} \hlkwb{<-} \hlstd{dat}\hlopt{$}\hlstd{SBP}\hlopt{/}\hlnum{250}
\hlstd{su} \hlkwb{<-} \hlkwd{c}\hlstd{(}\hlnum{0}\hlstd{,} \hlnum{1}\hlstd{)}
\hlstd{bo} \hlkwb{<-} \hlkwd{c}\hlstd{(}\hlnum{0}\hlstd{,} \hlnum{Inf}\hlstd{)}

\hlcom{### Multivariate models}
\hlstd{m1} \hlkwb{<-} \hlkwd{BoxCox}\hlstd{(Y1} \hlopt{~} \hlstd{D} \hlopt{*} \hlstd{(Age} \hlopt{+} \hlstd{Gender} \hlopt{+} \hlstd{Smoke),}
             \hlkwc{data} \hlstd{= dat,} \hlkwc{support} \hlstd{= su,} \hlkwc{bounds} \hlstd{= bo)}
\hlstd{m2} \hlkwb{<-} \hlkwd{BoxCox}\hlstd{(Y2} \hlopt{~} \hlstd{D} \hlopt{*} \hlstd{(Age} \hlopt{+} \hlstd{Gender} \hlopt{+} \hlstd{Smoke),}
             \hlkwc{data} \hlstd{= dat,} \hlkwc{support} \hlstd{= su,} \hlkwc{bounds} \hlstd{= bo)}
\hlstd{m_comb} \hlkwb{<-} \hlkwd{mmlt}\hlstd{(m1, m2,} \hlkwc{formula} \hlstd{=} \hlopt{~} \hlnum{1}\hlstd{,} \hlkwc{data} \hlstd{= dat)}
\hlstd{S} \hlkwb{<-} \hlkwd{coef}\hlstd{(m_comb,} \hlkwc{newdata} \hlstd{= dat[}\hlnum{1}\hlstd{,],} \hlkwc{type} \hlstd{=} \hlstr{"Sigma"}\hlstd{)}
\hlstd{cf} \hlkwb{<-} \hlkwd{coef}\hlstd{(m_comb)}
\hlstd{link} \hlkwb{<-} \hlstd{m_comb}\hlopt{$}\hlstd{marginals[[}\hlnum{1}\hlstd{]]}\hlopt{$}\hlstd{model}\hlopt{$}\hlstd{todistr}\hlopt{$}\hlstd{name}

\hlcom{### Indices}
\hlstd{d1idx} \hlkwb{<-} \hlkwd{which}\hlstd{(}\hlkwd{names}\hlstd{(cf)} \hlopt{==} \hlstr{"Y1.D"}\hlstd{)}
\hlstd{g1idx} \hlkwb{<-} \hlkwd{which}\hlstd{(}\hlkwd{grepl}\hlstd{(}\hlstr{"Y1.D:"}\hlstd{,}\hlkwd{names}\hlstd{(cf)))}
\hlstd{d2idx} \hlkwb{<-} \hlkwd{which}\hlstd{(}\hlkwd{names}\hlstd{(cf)} \hlopt{==} \hlstr{"Y2.D"}\hlstd{)}
\hlstd{g2idx} \hlkwb{<-} \hlkwd{which}\hlstd{(}\hlkwd{grepl}\hlstd{(}\hlstr{"Y2.D:"}\hlstd{,}\hlkwd{names}\hlstd{(cf)))}

\hlcom{### Covariate effect on the ROC curve}
\hlcom{### 40 year old female non-smoker}
\hlstd{x} \hlkwb{<-} \hlkwd{c}\hlstd{(}\hlnum{40}\hlstd{,} \hlnum{0}\hlstd{,} \hlnum{0}\hlstd{)}
\hlstd{lp1} \hlkwb{<-} \hlstd{(cf[d1idx]} \hlopt{+} \hlkwd{crossprod}\hlstd{(cf[g1idx], x))}
\hlstd{lp2} \hlkwb{<-} \hlstd{(cf[d2idx]} \hlopt{+} \hlkwd{crossprod}\hlstd{(cf[g2idx], x))} \hlopt{/} \hlkwd{sqrt}\hlstd{(S}\hlopt{$}\hlstd{diagonal[}\hlnum{2}\hlstd{])}

\hlcom{### Simulated parameters for 95 percent CIs}
\hlstd{V} \hlkwb{<-} \hlkwd{vcov}\hlstd{(m_comb)}
\hlstd{V} \hlkwb{<-} \hlstd{(V} \hlopt{+} \hlkwd{t}\hlstd{(V))} \hlopt{/} \hlnum{2}
\hlstd{nsim} \hlkwb{<-} \hlnum{1000}
\hlstd{P} \hlkwb{<-} \hlkwd{rmvnorm}\hlstd{(nsim,} \hlkwc{mean} \hlstd{= cf,} \hlkwc{sigma} \hlstd{= V)}
\hlstd{lp1s} \hlkwb{<-} \hlstd{P[,d1idx]} \hlopt{+} \hlkwd{crossprod}\hlstd{(}\hlkwd{t}\hlstd{(P[,g1idx]), x)}
\hlstd{lp2s} \hlkwb{<-} \hlstd{(P[,d2idx]} \hlopt{+} \hlkwd{crossprod}\hlstd{(}\hlkwd{t}\hlstd{(P[,g2idx]), x))} \hlopt{/} \hlkwd{sqrt}\hlstd{(S}\hlopt{$}\hlstd{diagonal[}\hlnum{2}\hlstd{])}
\hlstd{alpha} \hlkwb{<-} \hlnum{0.05}
\hlstd{qs} \hlkwb{<-} \hlkwd{c}\hlstd{(alpha}\hlopt{/}\hlnum{2}\hlstd{,} \hlnum{1}\hlopt{-}\hlstd{alpha}\hlopt{/}\hlnum{2}\hlstd{)}
\hlstd{cid1} \hlkwb{<-} \hlkwd{quantile}\hlstd{(lp1s, qs)}
\hlstd{cid2} \hlkwb{<-} \hlkwd{quantile}\hlstd{(lp2s, qs)}

\hlcom{### Marginal ROC curves}
\hlcom{## WHtR}
\hlstd{rm1} \hlkwb{<-} \hlkwd{ROC}\hlstd{(lp1,} \hlkwc{prob} \hlstd{= p,} \hlkwc{link} \hlstd{= link)}
\hlstd{lwr1} \hlkwb{<-} \hlkwd{ROC}\hlstd{(cid1[}\hlnum{1}\hlstd{],} \hlkwc{prob} \hlstd{= p,} \hlkwc{link} \hlstd{= link)}
\hlstd{upr1} \hlkwb{<-} \hlkwd{ROC}\hlstd{(cid1[}\hlnum{2}\hlstd{],} \hlkwc{prob} \hlstd{= p,} \hlkwc{link} \hlstd{= link)}

\hlcom{## SBP}
\hlstd{rm2} \hlkwb{<-} \hlkwd{ROC}\hlstd{(lp2,} \hlkwc{prob} \hlstd{= p,} \hlkwc{link} \hlstd{= link)}
\hlstd{lwr2} \hlkwb{<-} \hlkwd{ROC}\hlstd{(cid2[}\hlnum{1}\hlstd{],} \hlkwc{prob} \hlstd{= p,} \hlkwc{link} \hlstd{= link)}
\hlstd{upr2} \hlkwb{<-} \hlkwd{ROC}\hlstd{(cid2[}\hlnum{2}\hlstd{],} \hlkwc{prob} \hlstd{= p,} \hlkwc{link} \hlstd{= link)}

\hlkwd{plot}\hlstd{(p, rm1,} \hlkwc{type} \hlstd{=} \hlstr{"l"}\hlstd{,}
     \hlkwc{xlab} \hlstd{=} \hlstr{"1 - Specificity"}\hlstd{,} \hlkwc{ylab} \hlstd{=} \hlstr{"Sensitivity"}\hlstd{)}
\hlkwd{lines}\hlstd{(p, lwr1,} \hlkwc{lty} \hlstd{=} \hlnum{3}\hlstd{,} \hlkwc{col} \hlstd{=} \hlstr{"gray50"}\hlstd{)}
\hlkwd{lines}\hlstd{(p, upr1,} \hlkwc{lty} \hlstd{=} \hlnum{3}\hlstd{,} \hlkwc{col} \hlstd{=} \hlstr{"gray50"}\hlstd{)}
\hlkwd{lines}\hlstd{(p, rm2,} \hlkwc{lty} \hlstd{=} \hlnum{4}\hlstd{)}
\hlkwd{lines}\hlstd{(p, lwr2,} \hlkwc{lty} \hlstd{=} \hlnum{3}\hlstd{,} \hlkwc{col} \hlstd{=} \hlstr{"gray50"}\hlstd{)}
\hlkwd{lines}\hlstd{(p, upr2,} \hlkwc{lty} \hlstd{=} \hlnum{3}\hlstd{,} \hlkwc{col} \hlstd{=} \hlstr{"gray50"}\hlstd{)}
\hlkwd{legend}\hlstd{(}\hlstr{"bottomright"}\hlstd{,} \hlkwc{legend} \hlstd{=} \hlkwd{c}\hlstd{(}\hlstr{"WHtR"}\hlstd{,} \hlstr{"SBP"}\hlstd{),} \hlkwc{lty} \hlstd{=} \hlkwd{c}\hlstd{(}\hlnum{1}\hlstd{,} \hlnum{4}\hlstd{),} \hlkwc{bty} \hlstd{=} \hlstr{"n"}\hlstd{)}
\hlkwd{abline}\hlstd{(}\hlnum{0}\hlstd{,} \hlnum{1}\hlstd{,} \hlkwc{col} \hlstd{=} \hlstr{"lightgrey"}\hlstd{)}
\hlcom{### Marginal covariate-specific AUCs with 95 percent simulated intervals}
\hlcom{## WHtR}
\hlstd{(aucm1} \hlkwb{<-} \hlkwd{PI}\hlstd{(}\hlkwd{c}\hlstd{(lp1, cid1),} \hlkwc{link} \hlstd{= link))}
\hlcom{## SBP}
\hlstd{(aucm2} \hlkwb{<-} \hlkwd{PI}\hlstd{(}\hlkwd{c}\hlstd{(lp2, cid2),} \hlkwc{link} \hlstd{= link))}

\hlcom{### Comparison of correlated covariate-specific AUCs}
\hlstd{auc_dff} \hlkwb{<-} \hlkwd{PI}\hlstd{(lp1,} \hlkwc{link} \hlstd{= link)} \hlopt{-} \hlkwd{PI}\hlstd{(lp2,} \hlkwc{link} \hlstd{= link)}
\hlstd{dffs} \hlkwb{<-} \hlkwd{PI}\hlstd{(lp1s,} \hlkwc{link} \hlstd{= link)} \hlopt{-} \hlkwd{PI}\hlstd{(lp2s,} \hlkwc{link} \hlstd{= link)}
\hlstd{ci_dff} \hlkwb{<-} \hlkwd{quantile}\hlstd{(dffs, qs)}
\hlkwd{c}\hlstd{(auc_dff, ci_dff)}
\end{alltt}
\end{kframe}
\end{knitrout}

\end{document}